\documentclass{aa}  
\usepackage[pdftex]{color,xcolor}
\usepackage{graphicx}
\usepackage{amstext} 
\usepackage{amsmath} 
\usepackage{amssymb} 
\usepackage[figuresright]{rotating}
\usepackage{afterpage}
\usepackage{txfonts}

\newcommand{\dwnine}{DE0823$-$49}

\begin{document} 
   \title{Astrometric orbit of a low-mass companion to an ultracool dwarf\thanks{Based on observations made with ESO telescopes at the La Silla Paranal Observatory under programme IDs 086.C-0680, 088.C-0679, and 090.C-0786. Table 1 is only available in electronic form at the CDS via anonymous ftp to cdsarc.u-strasbg.fr (130.79.128.5) or via http://cdsweb.u-strasbg.fr/cgi-bin/qcat?J/A+A/.}} 
   
\author{J.~Sahlmann \inst{1}
		\and  P.~F.~Lazorenko \inst{2}
		\and D.~S\'egransan \inst{1}
		\and  E.~L.~Mart{\'{\i}}n \inst{3} 
		\and D.~Queloz\inst{1} 
	        \and M.~Mayor\inst{1} 
		\and S.~Udry\inst{1}}		
\institute{Observatoire de Gen\`eve, Universit\'e de Gen\`eve, 51 Chemin Des Maillettes, 1290 Sauverny, Switzerland\\
		\email{johannes.sahlmann@unige.ch}	
		\and
		Main Astronomical Observatory, National Academy of Sciences of the Ukraine, Zabolotnogo 27, 03680 Kyiv, Ukraine
		\and  
		Centro de Astrobiolog\'{\i}a (CSIC-INTA), Ctra. Ajalvir km 4, 28850 Torrej\'on de Ardoz, Madrid, Spain}

\date{Received 10 May 2013 / Accepted 13 June 2013} 
			
\abstract
{Little is known about the existence of extrasolar planets around ultracool dwarfs. Furthermore, binary stars with Sun-like primaries and very low-mass binaries composed of ultracool dwarfs show differences in the distributions of mass ratio and orbital separation that can be indicative of distinct formation mechanisms. Using {\small FORS2/VLT} optical imaging for high precision astrometry we are searching for planets and substellar objects around ultracool dwarfs to investigate their multiplicity properties for very low companion masses. Here we report astrometric measurements with an accuracy of {two tenths} of a milli-arcsecond over two years that reveal orbital motion of the {nearby L1.5 dwarf DENIS-P\,J082303.1-491201 located at $20.77 \pm 0.08$ pc} caused by an unseen companion that revolves about its host on an eccentric orbit in $246.4\pm1.4$ days. {We estimate the L1.5 dwarf to have $7.5\pm0.7$\% of the Sun's mass that implies a companion mass of $28\pm2$ Jupiter masses.} This new system has the smallest mass ratio ($0.36\pm0.02$) of known very low-mass binaries with characterised orbits. With this discovery we demonstrate {200} micro-arcsecond astrometry over an arc-minute field and over several years that is sufficient to discover sub-Jupiter mass planets around ultracool dwarfs. We also show that the achieved parallax accuracy of $<$\,0.4\,\% makes it possible to remove distance as a dominant source of uncertainty in the modelling of ultracool dwarfs.} 


\maketitle

\section{Introduction}
Brown dwarfs are abundant in the Galaxy \citep{Basri:2000vn} and appear to form like stars \citep{Luhman:2012vn} but are not massive enough to sustain Hydrogen fusion. Along with very low-mass stars they are referred to as ultracool dwarfs (spectral type M7 and later, \citealt{Kirkpatrick:2005vf}) that have masses in the range of approximately 0.01--0.1 times the mass of the Sun ($M_\sun$). They bridge the mass gap between planets and stars and provide us with a critical test of the understanding of planet and star formation. Observations of binary systems offer the opportunity for detailed studies and a handful of small separation ($\lesssim 1$ AU) ultracool binaries (i.e.\ both components are ultracool dwarfs) have been characterised \citep{Lane:2001le, Bouy:2004tg,Zapatero-Osorio:2004fu, Close:2005fk, Dupuy:2010lr} using spectroscopy \citep{Basri:1999kx}, eclipse photometry \citep{Stassun:2006pt}, astrometry \citep{Dahn:2008fk}, and gravitational microlensing \citep{Choi:2013fk}. Many systems were discovered using direct imaging, thus tend to be widely separated ($\gtrsim 1$ AU) and have components of similar brightness and nearly equal masses, i.e.\ large secondary/primary mass ratios $q=M_2/M_1$ (e.g.\ \citealt{Martin:1999uq,Bouy:2003kx,Close:2003kx}). Because of the associated long orbital periods {of} $\gtrsim10$ years, it is {tedious} to obtain dynamical mass and eccentricity constraints from orbital motion measurements of these wide systems (e.g.\ \citealt{Konopacky:2010dk}).\\
Ultracool binaries and binary stars with Sun-like primaries show differences in the distributions of mass ratio and orbital separation \citep{Burgasser:2007ix} that may be signatures of distinct formation mechanisms \citep{Whitworth:2007pi, Thies:2008ve, Goodwin:2013fk}. Most of the known systems have large mass ratios and are widely separated, which in part is attributable to sensitivity limitations of the respective observing techniques. Consequently, the occurrence and configurations of small-separation ultracool binaries with small mass ratios are uncertain. This includes extrasolar planets that are common around Sun-like stars \citep{Mayor:2011fj} but have so far not been found around brown dwarfs. Two very low-mass stars are known to host Earth-mass planets \citep{Kubas:2010fk, Muirhead:2012fk} and radial velocity and direct imaging {surveys} could exclude a large population of giant planets $>\!1$ Jupiter mass ($M_J$) in close orbits $<$\,0.05 AU and at wide separations $\gtrsim$\,2 AU around ultracool dwarfs \citep{Blake:2010lr, Stumpf:2010lr}.\\
Astrometric measurements determine the positions of stars in the plane of the sky and make it possible to discover and characterise multiple stellar systems through the detection of orbital motion. Low-mass systems containing brown dwarfs and extrasolar planets are difficult to study in this way because they are faint and the signatures of orbital motion have typical amplitudes smaller than one milli-arcsecond (mas) \citep{Black:1982kx}, which is a challenging figure for currently available instruments. However, astrometric measurements of 50-100 micro-arcseconds ($\mu$as) precision were demonstrated by \cite{Lazorenko:2006qf,Lazorenko:2007ul,Lazorenko:2009ph,Lazorenko:2011lr} using ground-based optical imaging with an 8~m class telescope. To exploit this capability, we have initiated a planet search survey around ultracool dwarfs that will be described in detail in a forthcoming paper. Here, we report the first result of this survey: the detection and characterisation of a low-mass companion to an ultracool dwarf.

\section{Observations and data reduction}
We observed the nearby ultracool dwarf \object{DENIS-P\,J082303.1-491201} (\citealt{Phan-Bao:2008fr}, spectral type L1.5, hereafter \dwnine) with the {\small FORS2} \citep{Appenzeller:1998lr} seeing-limited optical camera of the Very Large Telescope on 14 epochs between October 2010 and January 2013. The observations were separated by typically one month during the seasonal windows and at each epoch we obtained 25 consecutive $I$-band images of the target field located close to the southern Galactic plane and containing several hundred reference stars with brightnesses of approximately 17th - 22nd magnitude, thus similar to \dwnine~having an $I$-band magnitude of $m_I=17.1$. The position of the target relative to the local grid of reference stars was determined using an improved version of the methods described in \cite{Lazorenko:2009ph, Lazorenko:2011lr}. The dense stellar field is used to correct for atmospheric image motion, optical distortions introduced by the telescope and camera system, and systematic displacement errors at the level of one-thousandth of a detector pixel. One central element of the method relies on averaging the turbulence occurring in the Earth's atmosphere above the telescope over its 8.2~m aperture. 
{The relative positions of reference stars are free to vary between frames due to proper motion and parallax. Additionally, they are affected by differential chromatic refraction that displaces 'blue' stars towards zenith and 'red' stars in the opposite direction. This displacement is typically 1--10~mas and can be computed using a star's colour index \citep[Eq. 19]{Lazorenko:2006qf}, however, a better correction can be obtained by modelling it with a free parameter \citep{Lazorenko:2011lr}.} The typical epoch precision of the astrometric observations was 0.1--0.2 mas, comparable to our observations of the ultracool dwarf \object{VB\,10} \citep{Lazorenko:2011lr}, and allowed us to accurately monitor the sky-projected motion of \dwnine. After accounting for proper and parallactic motion, an additional signal was detected with an amplitude much larger than the measurement precision. The astrometric data were therefore searched for evidence of orbital motion and {processed} with a Bayesian analysis package consisting of a genetic algorithm followed by a Markov Chain Monte Carlo analysis. 

\section{Orbit adjustment and parameter estimation}
The target's astrometric motion is modelled with the prescription Eq. \ref{axmodel}, where $\alpha^{\star}_m$ and $\delta_m$ are the astrometric measurements {relative to the grid of reference stars} in right ascension ({\small RA}) and declination  ({\small DEC}), respectively, in frame $m$ taken at time $t_m$ \citep{Lazorenko:2011lr, Sahlmann:2011fk} 
\begin{equation}\label{axmodel}
\begin{array}{ll@{\hspace{1mm}}l@{\hspace{1mm}}l}
\!\!\alpha^{\star}_{m} =\!\!\!\!\!\!& \Delta \alpha^{\star}_0 + \mu_{\alpha^\star} \, t_m + \varpi \, \Pi_{\alpha,m} &- \rho\, f_{1,x,m}-  d \,f_{2,x,m} &+ (B \, X_m + G \, Y_m)\\
\!\delta_{m} = \!\!\!\!\!\!& \underbrace{\Delta \delta_0 + \,\mu_\delta      \,  \;                      t_m \;+ \varpi \, \Pi_{\delta,m}}_{\mbox{Standard model} }  &\underbrace{+ \rho \,f_{1,y,m}+  d \,f_{2,y,m}}_{\mbox{Refraction}} &\underbrace{+ (A \, X_m + F \, Y_m)}_{\mbox{Orbital motion}}.
\end{array}
\end{equation}
The standard astrometric model consists of coordinate offsets $\Delta\alpha^{\star}_0, \Delta\delta_0$, the target's proper motions $\mu_{\alpha^\star}, \mu_\delta$, and the parallactic motion expressed as the product of relative parallax $\varpi$ and the parallax factors $\Pi_\alpha, \Pi_\delta$. {This parallax is not absolute because it is measured relative to the reference stars that are not located at infinite distances, which makes a parallax correction necessary (Sect.~\ref{sec:parcor}). We require reference stars to have zero parallax on average and about half of the reference stars have therefore negative parallaxes. }Differential chromatic refraction is modelled with the parameters $\rho$ and $d$ whose values depend on the star's colour. {The parameter $d$ is necessary because the observations were obtained with the telescope's longitudinal atmospheric dispersion compensation mechanism \citep{Avila} that improves the image quality and has one degree of freedom, which is the average zenith angle $z_L$ of an observation.} The factors
\begin{equation}\label{eq:x}
\begin{array}{ll@{\hspace{0mm}}l}
f_{1,x,m} &= f_{3,m}\,\, &\tan z_m  \sin \gamma_m\\
f_{2,x,m} &= &\tan z_{{\rm L},m} \sin \gamma_m\\
f_{1,y,m} &= f_{3,m}\,\, &\tan z_m  \cos \gamma_m\\
f_{2,y,m} &= &\tan z_{{\rm L},m} \cos \gamma_m,
\end{array}
\end{equation}
where $z$ is the zenith angle and $\gamma$ is the angle between the direction to zenith and the $y$-axis (declination), depend on ambient temperature $T_a$ in degree Celsius and pressure $P_a$ in hPa 
\begin{equation}\label{eq:xx}
f_{3,m} = \left(1-\frac{T_{a,m}-11}{273+11}\right)\left(1+\frac{P_{a,m}-744}{744}\right).
\end{equation}
Finally, orbital motion is modelled as a Keplerian two-body system, where $A,B,F,G$ are the Thiele-Innes constants (that map the barycentric orbit semimajor axis $a_1$, the argument of periastron $\omega$, the inclination $i$, and the ascending node $\Omega$) and $X_m, Y_m$ are the elliptical-rectangular coordinates that depend on eccentricity $e$ and eccentric anomaly $E$. The time-dependent eccentric anomaly $E_m\left(t_m;\,e,M_{0},P\right)$ is also a function of the mean anomaly $M_{0}$ and the orbital period $P$. There are thus 14 free parameters.
\addtocounter{table}{1}
{The astrometry data and the parameters necessary to apply the model function Eq.~\ref{axmodel} are given in Table 1, available at the CDS, that contains the following information for the 281 individual frames. Column 1 lists the epoch number, Column 2 gives the Modified Julian Date of the observation, Columns 3-6 give the relative astrometry and uncertainties ($\alpha^{\star}$, $\sigma_{\alpha^{\star}}$, $\delta$, $\sigma_\delta$), Columns 7 and 8 list the parallax factors, Columns 9-12 give the factors $f_{1,x,m}$, $f_{2,x,m}$, $f_{1,y,m}$, $f_{2,y,m}$, and Column 13 lists the covariance.} 

\subsection{Genetic algorithm}
The genetic algorithm is designed to efficiently, yet comprehensively, probe a large parameter space and to identify the globally best set of parameters. We divided the 14 free parameters into the two groups of linear and non-linear parameters on the basis of their appearance in Eq.~\ref{axmodel}. The genetic part of the algorithm probes the non-linear parameters of the Keplerian equations, namely the eccentricity, the mean anomaly, and the period, whereas the remaining 11 linear parameters ($\Delta\alpha^{\star}_0, \Delta\delta_0, \mu_{\alpha^\star}, \mu_\delta, \varpi, A,B,F,G,\rho, d$) are obtained using a standard least-square fit. Initial guesses for the orbital period are obtained from a generalised Lomb-Scargle periodogram and the set of best solutions is retrieved using the Bayesian formalism by {maxi}mising $\log{(\rm Posterior)}= \log{(\rm Likelihood \times Priors)}$. The list of priors used to build the merit function is given in Table~\ref{tab-ga-params}.
\begin{table}[h!]
\center
\caption{Parameters probed by the genetic algorithm}             
\label{tab-ga-params}      
\begin{tabular}{lccl}     
\hline    
\hline
Param. & Unit& Prior& Description\\
\hline                  
$P$ & day&Jeffreys&Period \\
$M_{0}$ &deg&Uniform& Mean anomaly at ref. date \\
$e$&$\cdots$&Uniform& Eccentricity\\ 
\hline                  
\end{tabular}
\end{table}

\subsection{Markov Chain Monte Carlo}
Posterior sampling using the genetic algorithm is efficient to identify the best solution but it does not lead to a statistically reliable sample that can be used to obtain robust parameter distributions and determine confidence intervals. We therefore sample the posterior distributions using a Markov Chain Monte Carlo analysis with Metropolis-Hastings ({\small MCMC}). {\small MCMC} posterior sampling \citep{Andrieu:2008qy} is a commonly implemented method, for instance in exoplanet research \citep{Gregory:2005kx, Gregory:2005vn,Collier-Cameron:2007ys, Pollacco:2008zr}. We start the {\small MCMC} by drawing several chains (typically 5) from the last chromosome generation of the genetic algorithm to compare their convergence. The step sizes are derived according to the r.m.s. of 95\% of the last population obtained by the genetic algorithm. The astrometric model is no longer split into linear and non-linear parameters, instead all 14 parameters are probed with {the} different set of priors listed in Table~\ref{tab-mcmc-params}. Two additional nuisance parameters are added to take into account potential signals that are not accounted for by the model \citep{Pollacco:2008zr}. Those terms affect the likelihood as well as the priors. We use uniform priors for all parameters except for ${P}$, ${a_{1}}$ and ${\varpi}$, for which modified Jeffreys priors are preferred \citep{Gregory:2005vn}. We use $\sqrt{e}\cos{\omega}$ and $\sqrt{e}\sin{\omega}$ as free parameters that translate into a uniform prior in eccentricity \citep{Anderson:2011uq}. The mean longitude $\lambda_{0} = M_0 + \omega$ is preferred over the mean anomaly or the date of periastron passage because it is not degenerate at low eccentricities. A large number of {\small MCMC} iterations allows us to retrieve a statistically reliable posterior distribution and the marginal parameter distributions.
\begin{table}[h!]
\caption{Parameters probed by the {\small MCMC}.\vspace{-0.8cm}}             
\label{tab-mcmc-params}      
\center
\begin{tabular}{cccl}     
\hline
\hline    
Parameter & Unit& Prior& Description\\
\hline                  
$\Delta\alpha^{\star}_0, \Delta\delta_0$& mas&Uniform&Coordinate offsets\\
$\mu_{\alpha^\star}, \mu_\delta$&mas/yr&Uniform&Proper motions\\
$\varpi $&mas&Jeffreys&Relative parallax\\
$s_{\alpha},s_{\delta}$  & mas &Uniform&Nuisance parameters\\
$P$ & day&Jeffreys&Period \\
$a_{1}$ & mas&Jeffreys&Semimajor axis \\
$\lambda_{0} $ &deg&Uniform& Mean longitude at ref. date\\
$\sqrt{e}\,\cos{\omega}$&$\cdots$&Uniform&Ecc. and arg. of periastron\\ 
$\sqrt{e}\,\sin{\omega}$  &$\cdots$&Uniform&Ecc. and arg. of periastron\\
$\Omega$  &deg&Uniform&Long. of ascending node\\
$i$  &deg&Uniform&Inclination\\
$\rho$  &mas&Uniform& Refraction coefficient\\
$d$  &mas&Uniform& Refraction coefficient\\
\hline                  
\end{tabular}
\end{table}

\subsection{MCMC results}
We run the {\small MCMC} for $10^{7}$ iterations and the statistics are derived on the last $7.5\cdot10^{6}$ elements. A characteristic of the reduction procedure is that astrometric measurements taken within one epoch are correlated and the corresponding covariance matrix has non-zero off-diagonal entries. In practice, however, their effect is taken into account by the nuisance parameters and we therefore work with a diagonal covariance matrix (see Appendix~\ref{sec:cov}). The marginal parameter distributions corresponding to the astrometric data of \dwnine~is shown in Fig.~\ref{fig:PDF1Param}. Most parameters show a nearly Gaussian distribution. Joint marginal parameter distributions are represented in Fig.~\ref{fig:PDF2Param} and used to identify correlations between different sets of parameters. The parameters show no significant correlation with the exception of a weak correlation between the parallax and the barycentric orbit semi-major axis. The median values and the 1-$\sigma$ confidence intervals of the adjusted parameters are given in Table \ref{table:1}. The coordinate offsets $\Delta\alpha^{\star}_0, \Delta\delta_0$ are relative to the target's position at the reference date. An additional noise (nuisance parameter) of 0.16 mas and 0.12 mas in {\small RA} and {\small DEC}, respectively, is required for the model adjustment. {We separately list the r.m.s. of the 281 individual frame residuals ($\sigma_{O - C}$) and the r.m.s. of the 14 epoch-averaged residuals ($\sigma_{O - C, Epoch} =0.330$ mas). The latter is} slightly higher than the mean epoch uncertainty of 0.188 mas. The excess noise originates primarily in the {\small RA} measurements (0.410 mas r.m.s. compared to 0.221 mas r.m.s. of {\small DEC} measurements) and is attributable to the presence of a background star located 1\,--\,0.7$\arcsec$ west on the target's projected trajectory. The angular separation of both objects decreases with time to the point that it degrades the photocentre measurement especially in the second and third observation season (Fig.~\ref{fig:axt},a,b).
This is also reflected in the reduced chi-squared value $\chi^2_\mathrm{red}=3.40$ that significantly exceeds unity. For {\small DEC} measurements, $\chi^2_\mathrm{red, \delta}=1.62$ is better and comparable to  $\chi^2_\mathrm{red,ref}=1.72\pm 0.48$ (r.m.s.) for bright reference stars. The fit quality of \dwnine~in {\small DEC} is thus equal to that of distant reference stars.
\onlfig{
\begin{figure*}[!h] 
\hbox to \textwidth{
\parbox{0.33\textwidth}{
\includegraphics[angle=0,width=0.32\textwidth,origin=br]{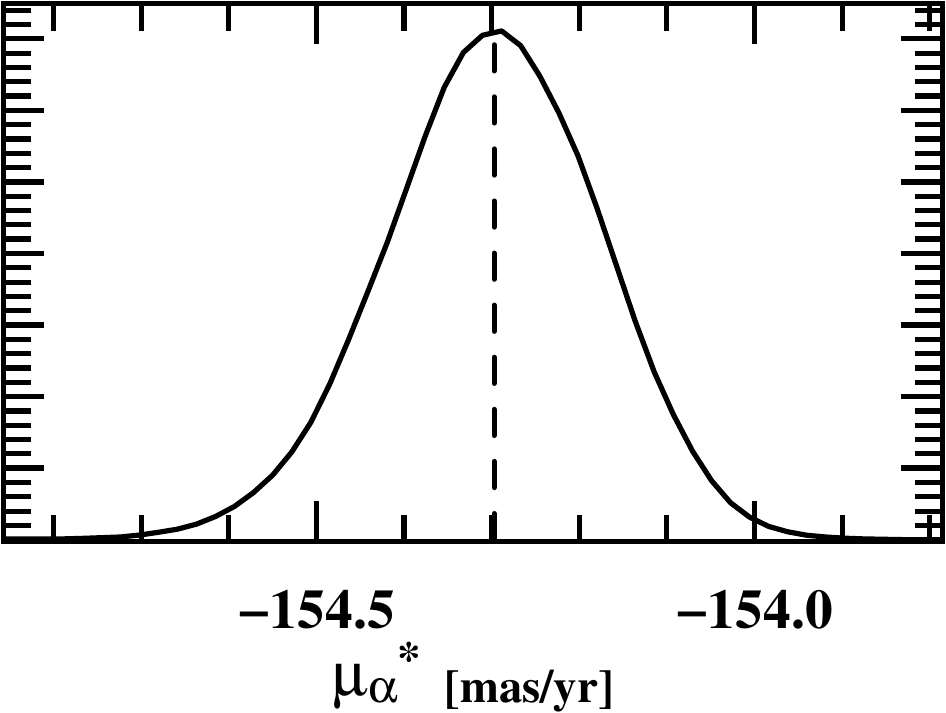}}
\hfil
\parbox{0.33\textwidth}{
\includegraphics[angle=0,width=0.32\textwidth,origin=br]{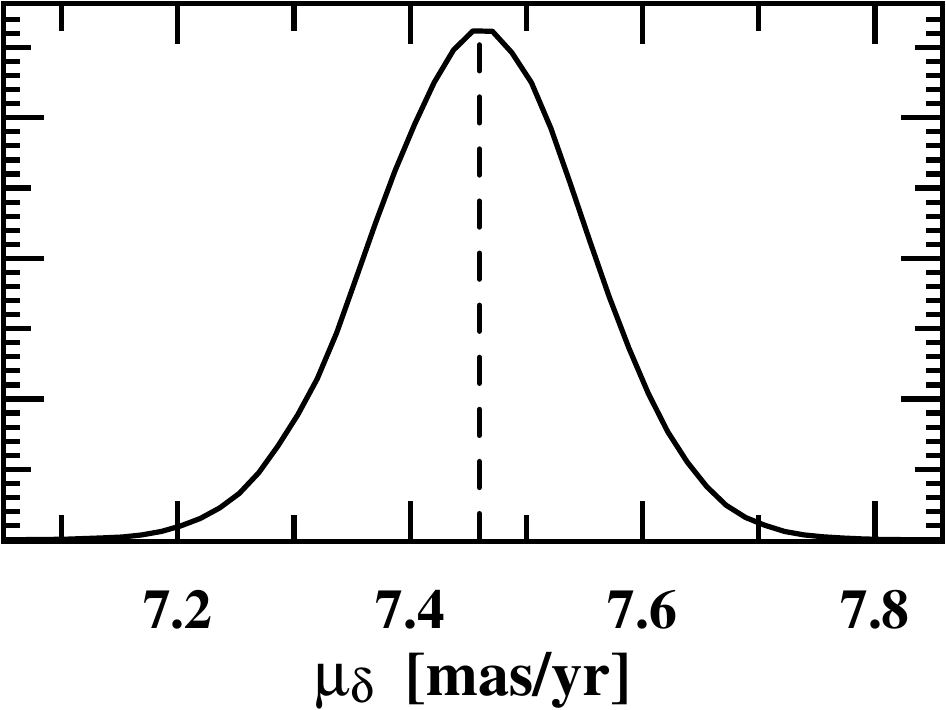}}
\hfil
\parbox{0.33\textwidth}{
\includegraphics[angle=0,width=0.32\textwidth,origin=br]{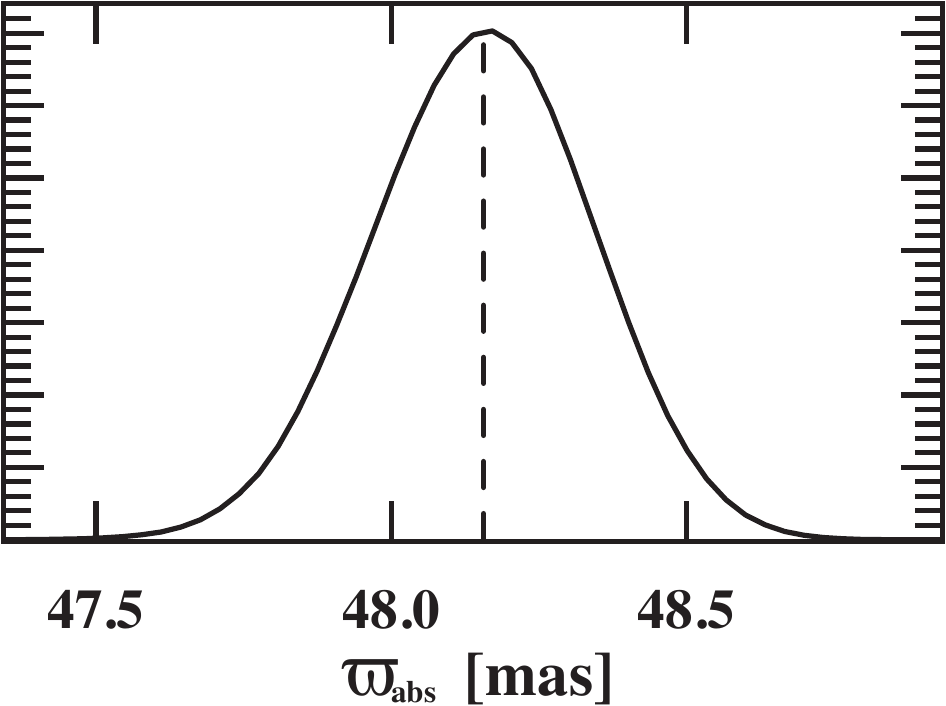}}}
\vspace{3truemm}\par\noindent
\hbox to \textwidth{
\parbox{0.25\textwidth}{
\includegraphics[angle=0,width=0.24\textwidth,origin=br]{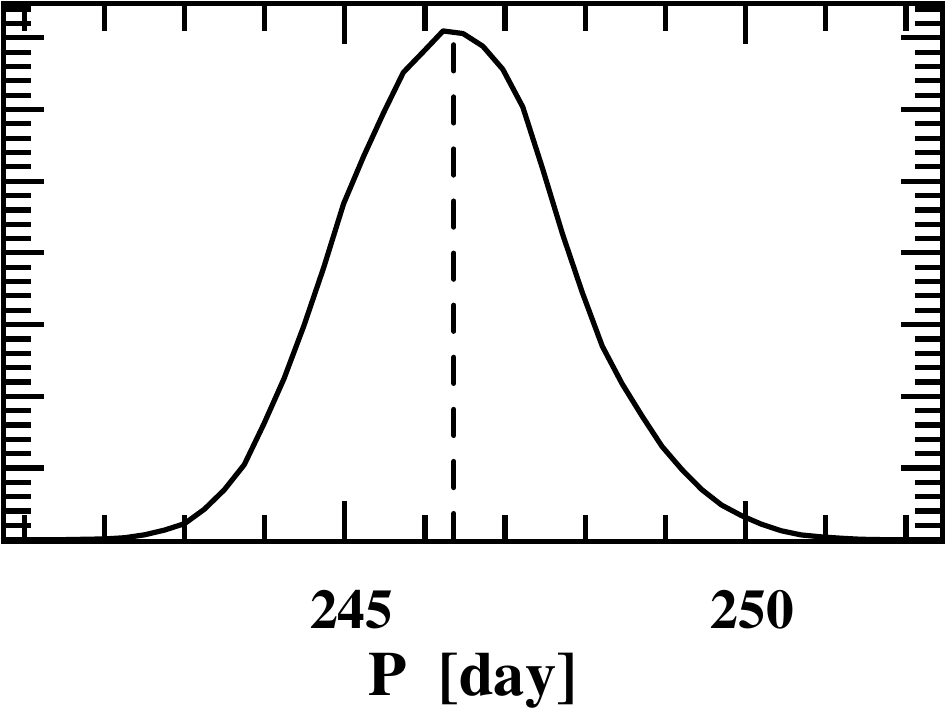}}
\hfil
\parbox{0.25\textwidth}{
\includegraphics[angle=0,width=0.24\textwidth,origin=br]{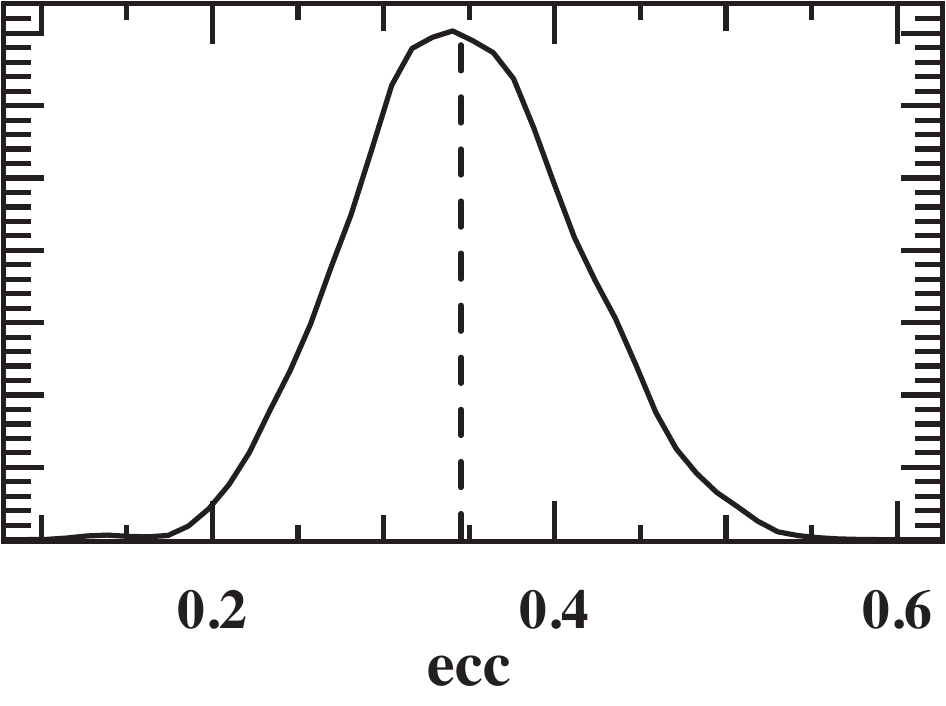}}
\hfil
\parbox{0.25\textwidth}{
\includegraphics[angle=0,width=0.24\textwidth,origin=br]{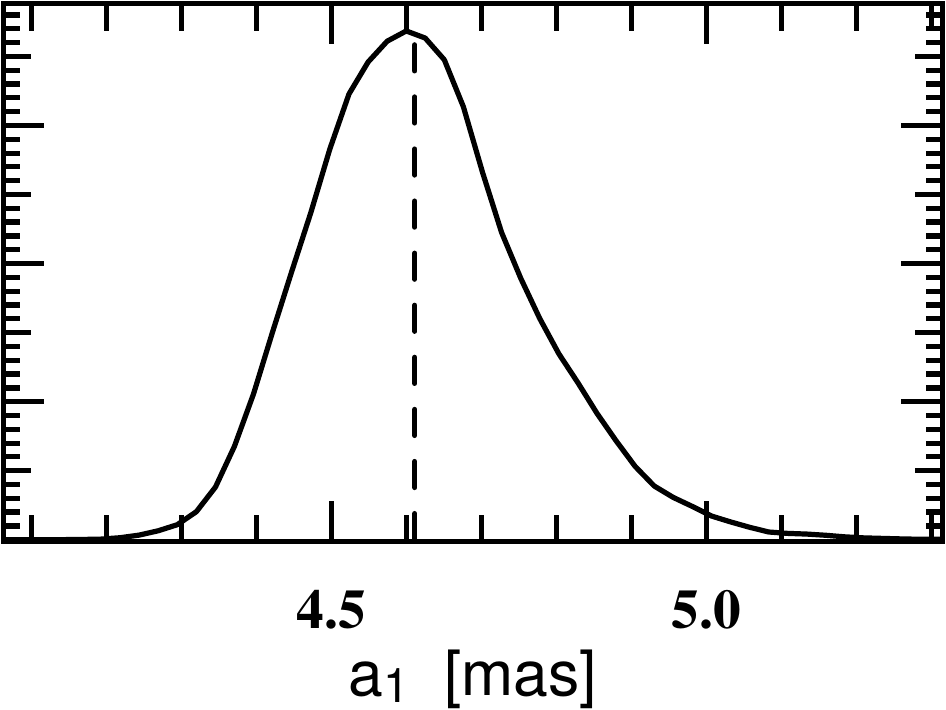}}
\hfil
\parbox{0.25\textwidth}{
\includegraphics[angle=0,width=0.24\textwidth,origin=br]{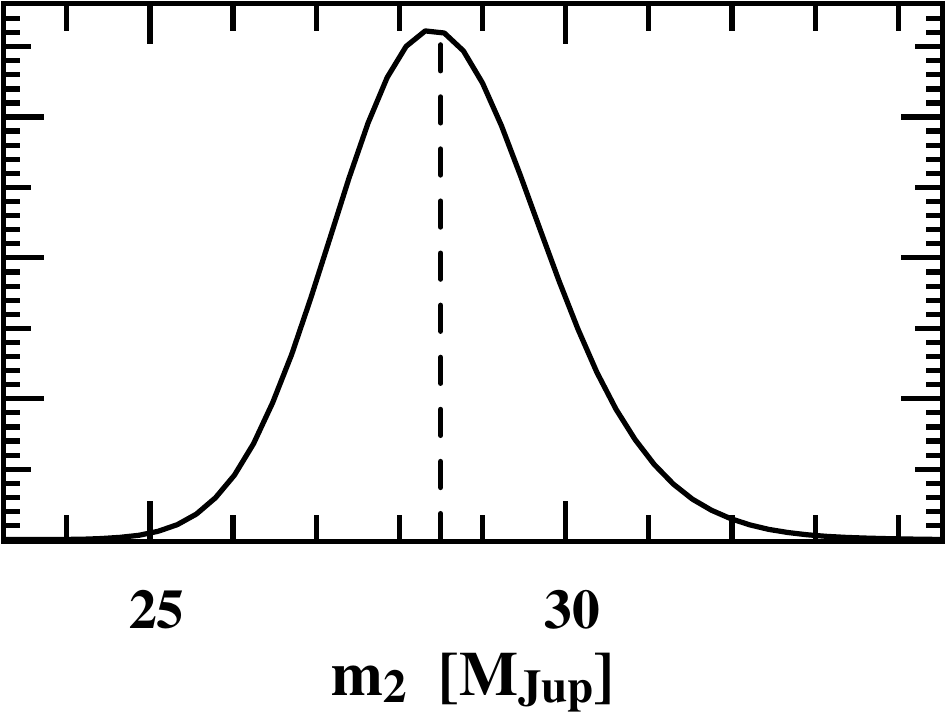}}}
\vspace{3truemm}\par\noindent
\hbox to \textwidth{
\parbox{0.25\textwidth}{
\includegraphics[angle=0,width=0.24\textwidth,origin=br]{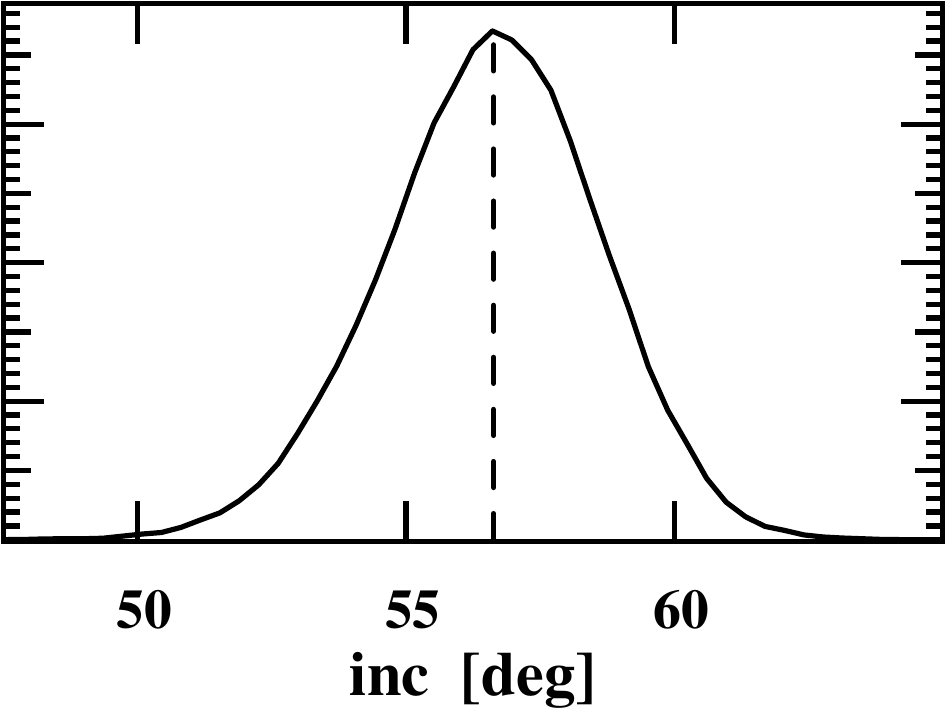}}
\hfil
\parbox{0.25\textwidth}{
\includegraphics[angle=0,width=0.24\textwidth,origin=br]{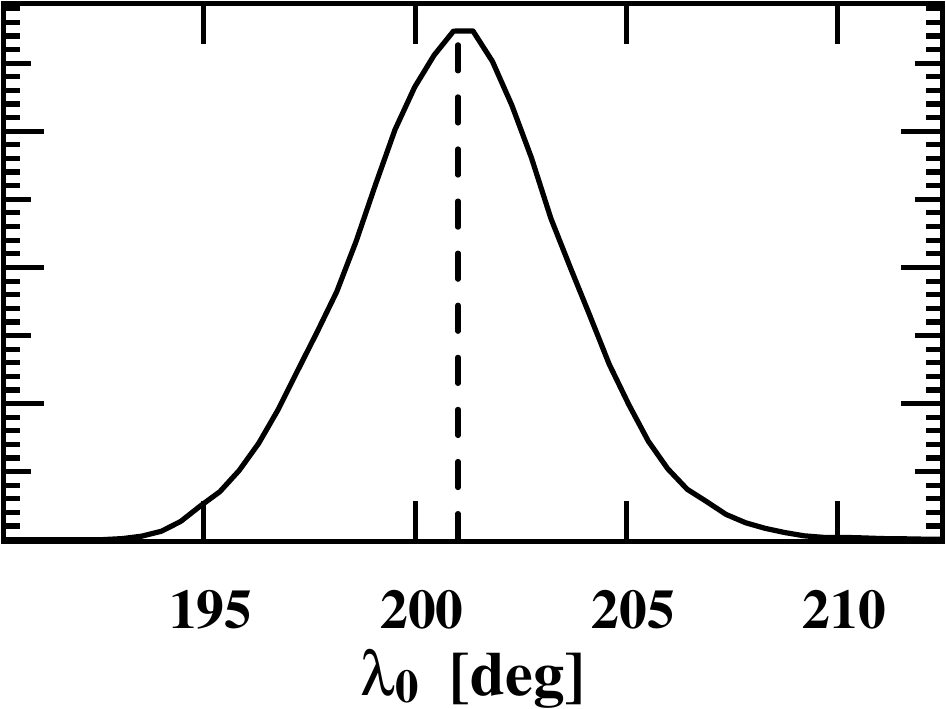}}
\hfil
\parbox{0.25\textwidth}{
\includegraphics[angle=0,width=0.24\textwidth,origin=br]{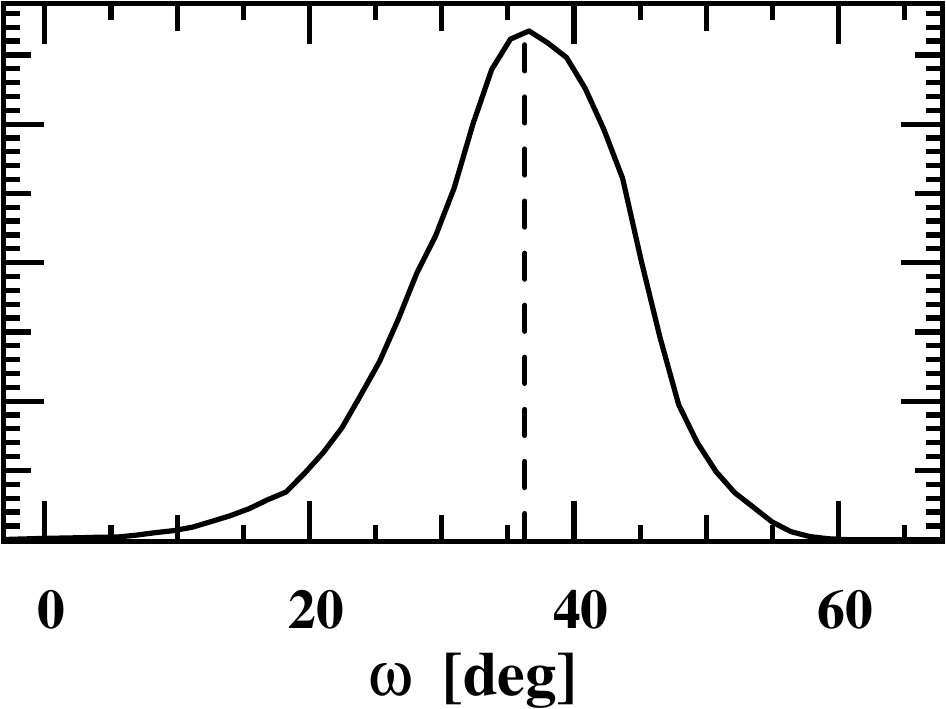}}
\hfil
\parbox{0.25\textwidth}{
\includegraphics[angle=0,width=0.24\textwidth,origin=br]{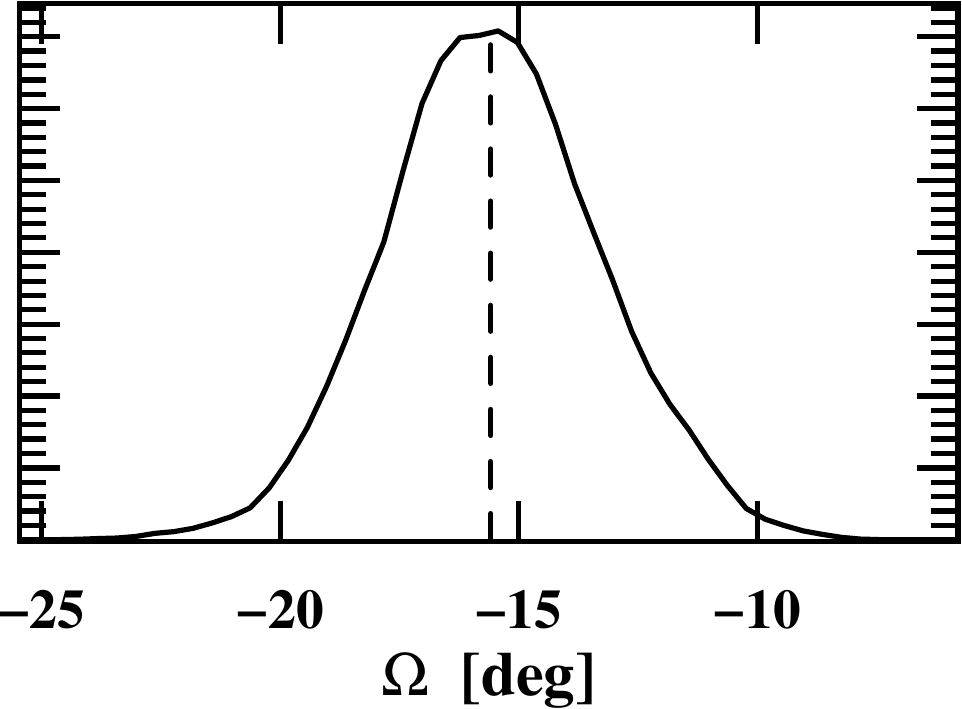}}}
\vspace{3truemm}\par\noindent
\hbox to \textwidth{
\parbox{0.33\textwidth}{
\includegraphics[angle=0,width=0.32\textwidth,origin=br]{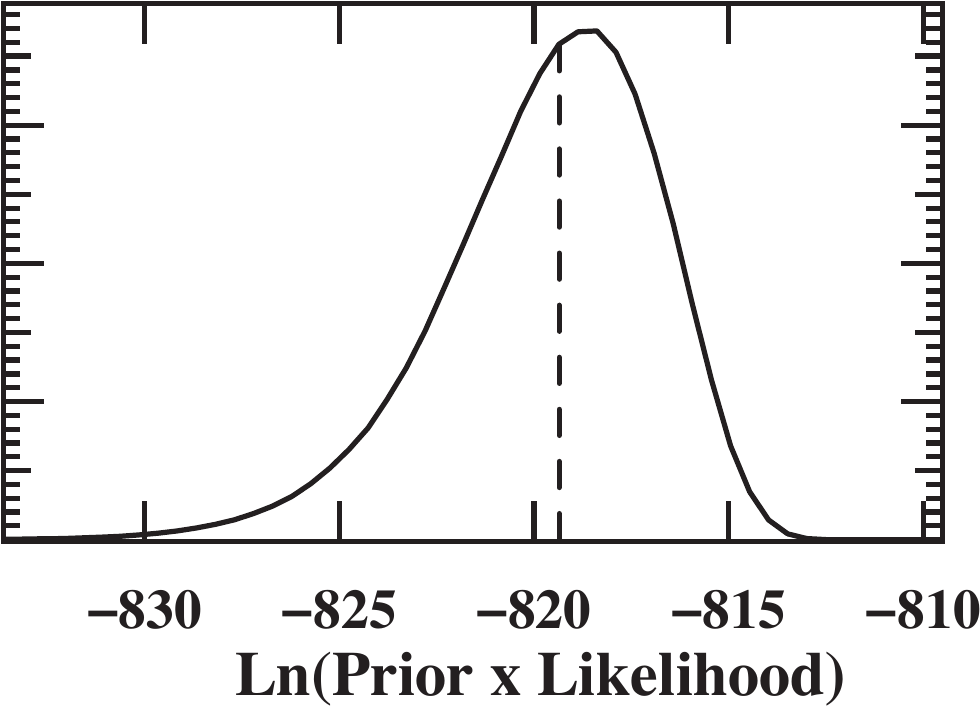}}
\hfil
\parbox{0.33\textwidth}{
\includegraphics[angle=0,width=0.32\textwidth,origin=br]{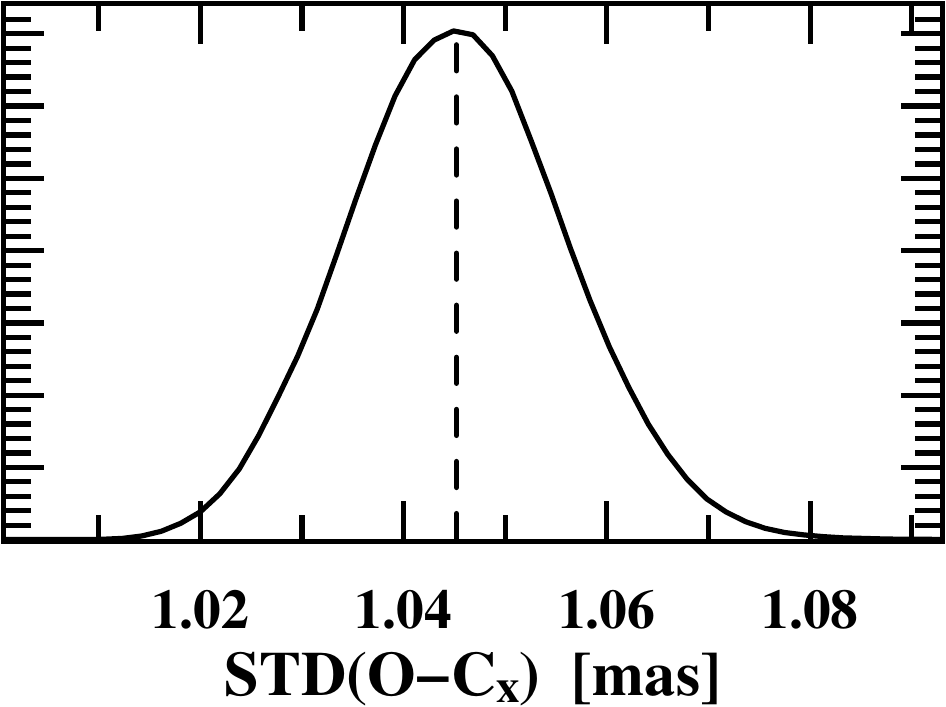}}
\hfil
\parbox{0.33\textwidth}{
\includegraphics[angle=0,width=0.32\textwidth,origin=br]{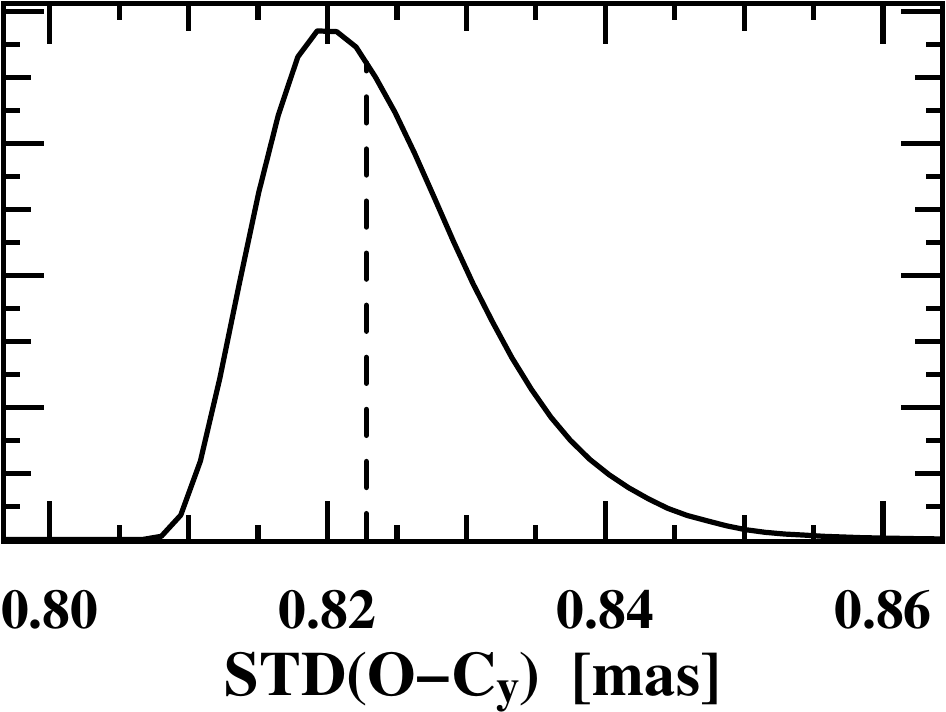}}}
\caption{Marginal parameter distributions for \dwnine~obtained from $7.5\cdot10^{6}$ MCMC iterations. In each panel, the dashed line indicates the median value of the respective parameter identified by the x-coordinate lable. Y-coordinate units are arbitrary and indicate relative occurrence.} 
\label{fig:PDF1Param}
\end{figure*}
}
\onlfig{
\begin{figure*}[!h] 
\centering
\vspace{2truemm}\par\noindent
\hbox to \textwidth{
\parbox{0.45\textwidth}{
\includegraphics[angle=0,width=0.44\textwidth,origin=br]{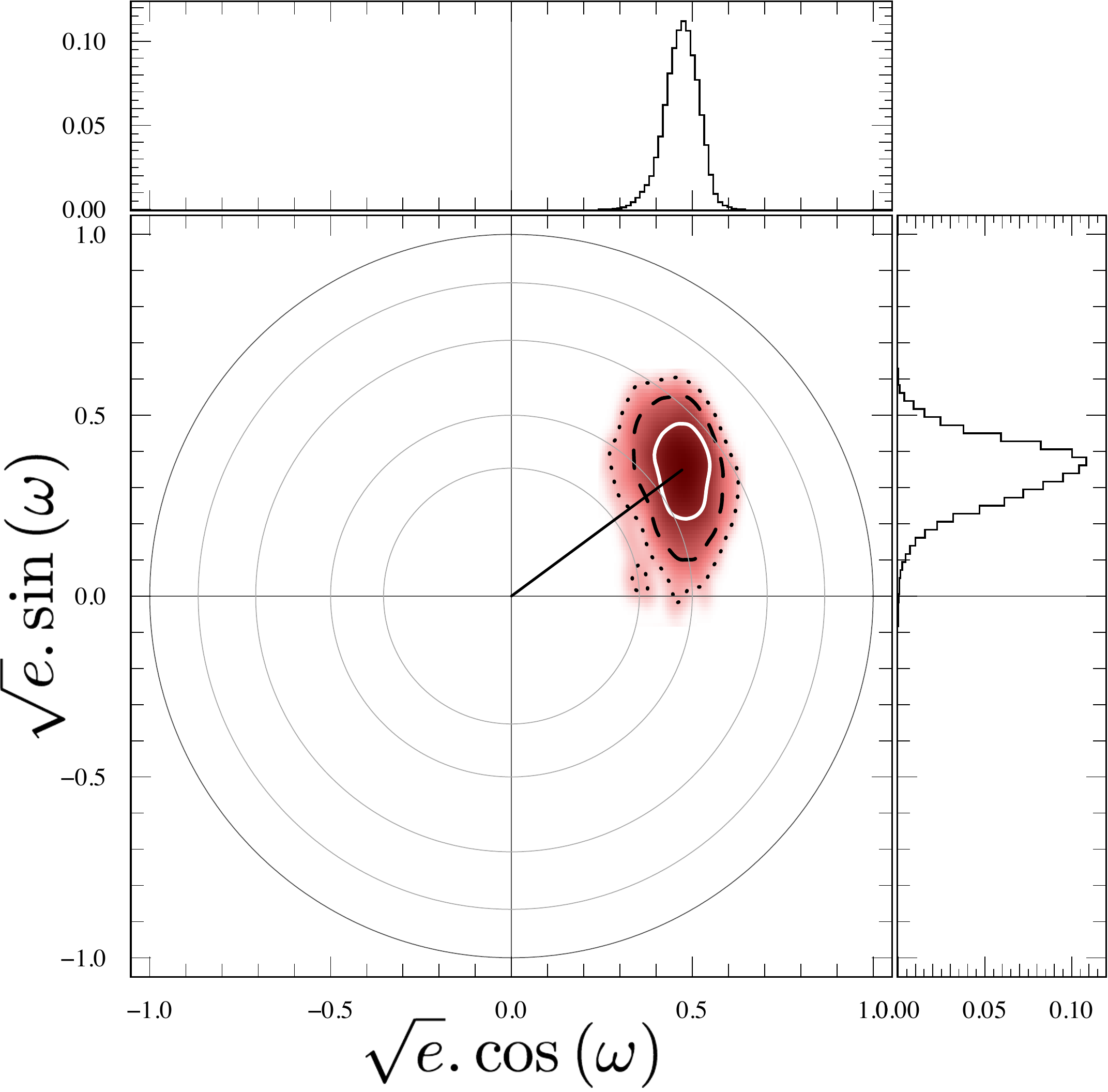}}
\hfil
\parbox{0.45\textwidth}{
\includegraphics[angle=0,width=0.44\textwidth,origin=br]{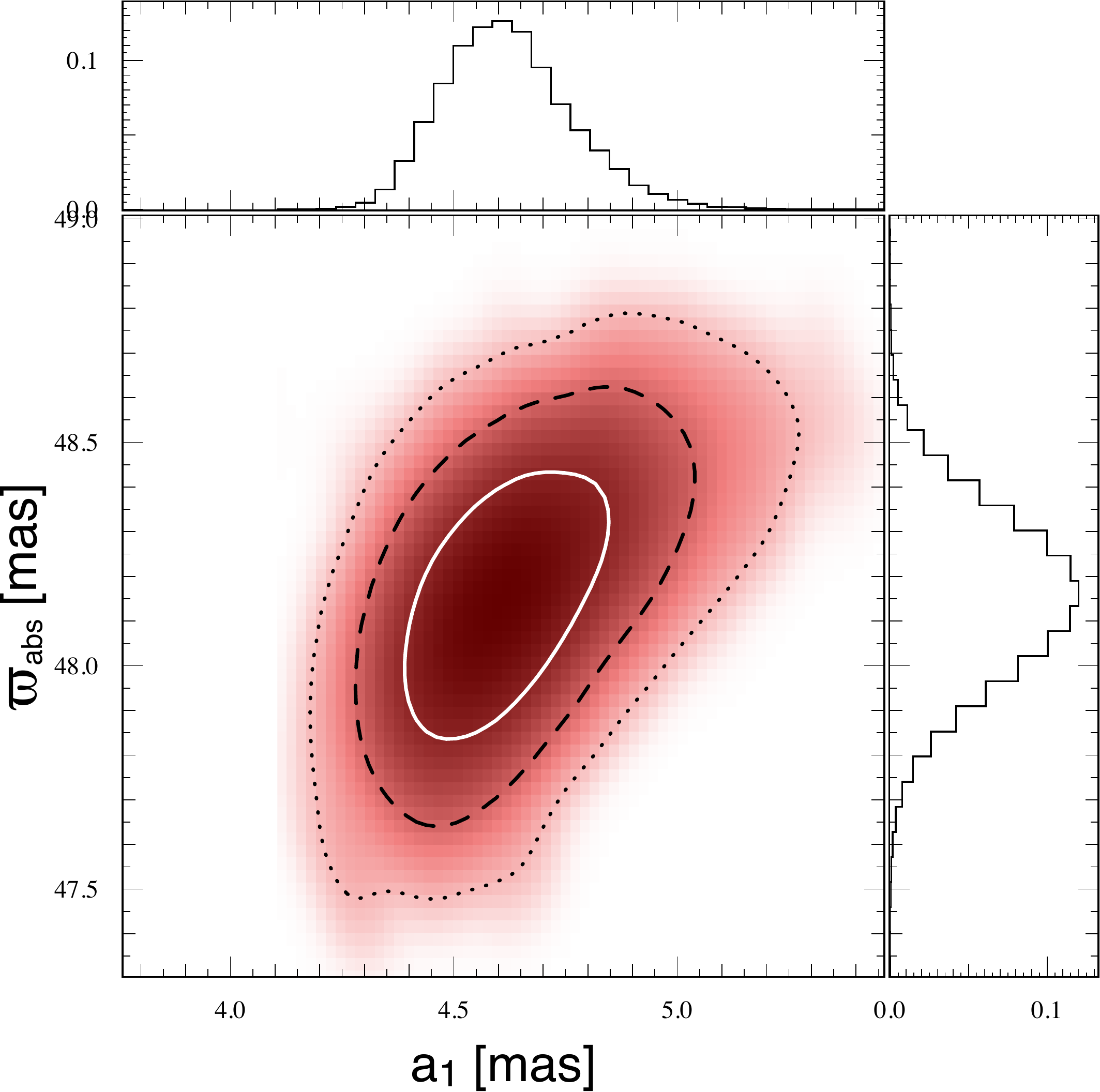}}}
\vspace{1truemm}\par\noindent
\hbox to \textwidth{
\parbox{0.45\textwidth}{
\includegraphics[angle=0,width=0.44\textwidth,origin=br]{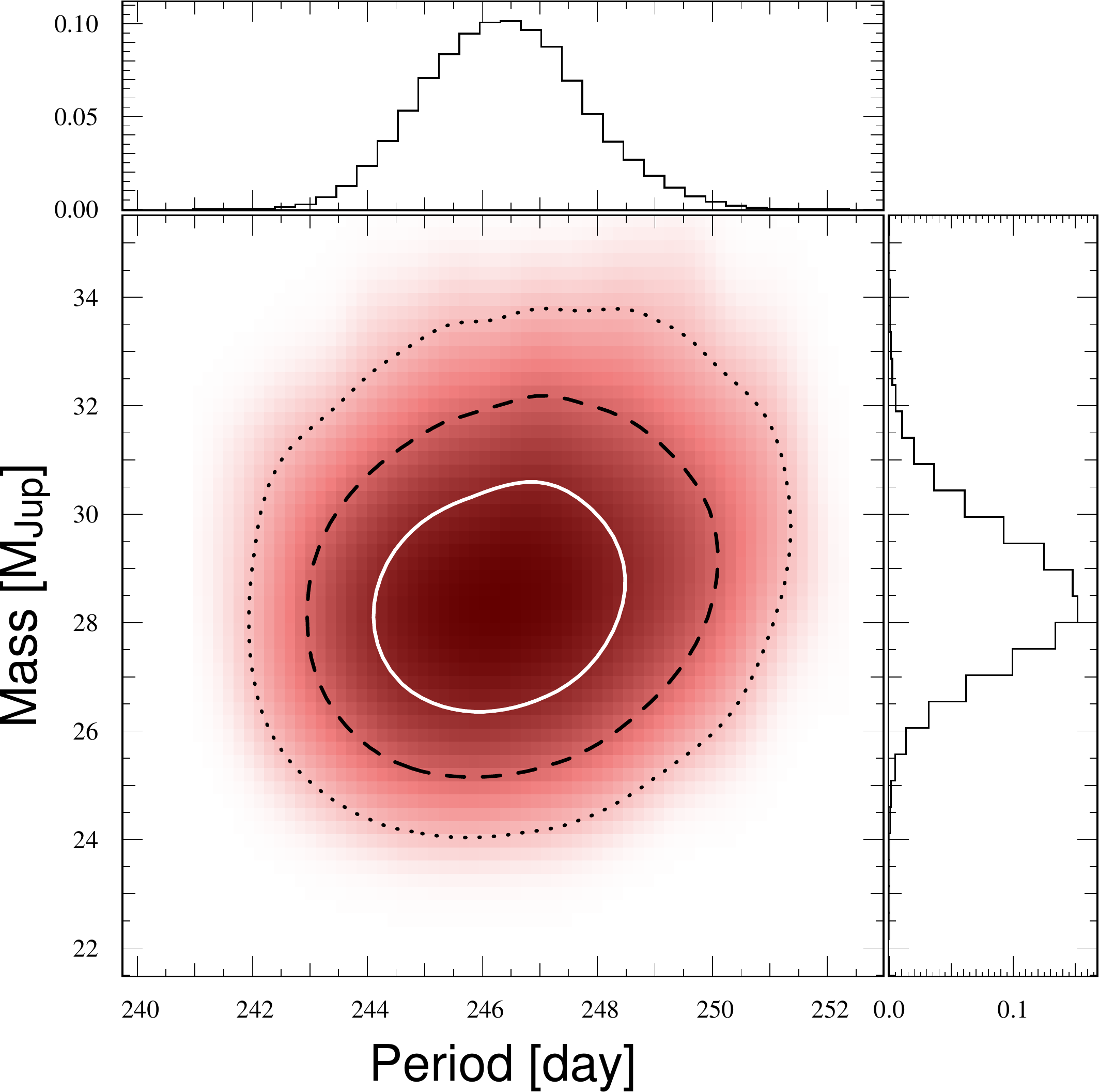}}
\hfil
\parbox{0.45\textwidth}{
\includegraphics[angle=0,width=0.44\textwidth,origin=br]{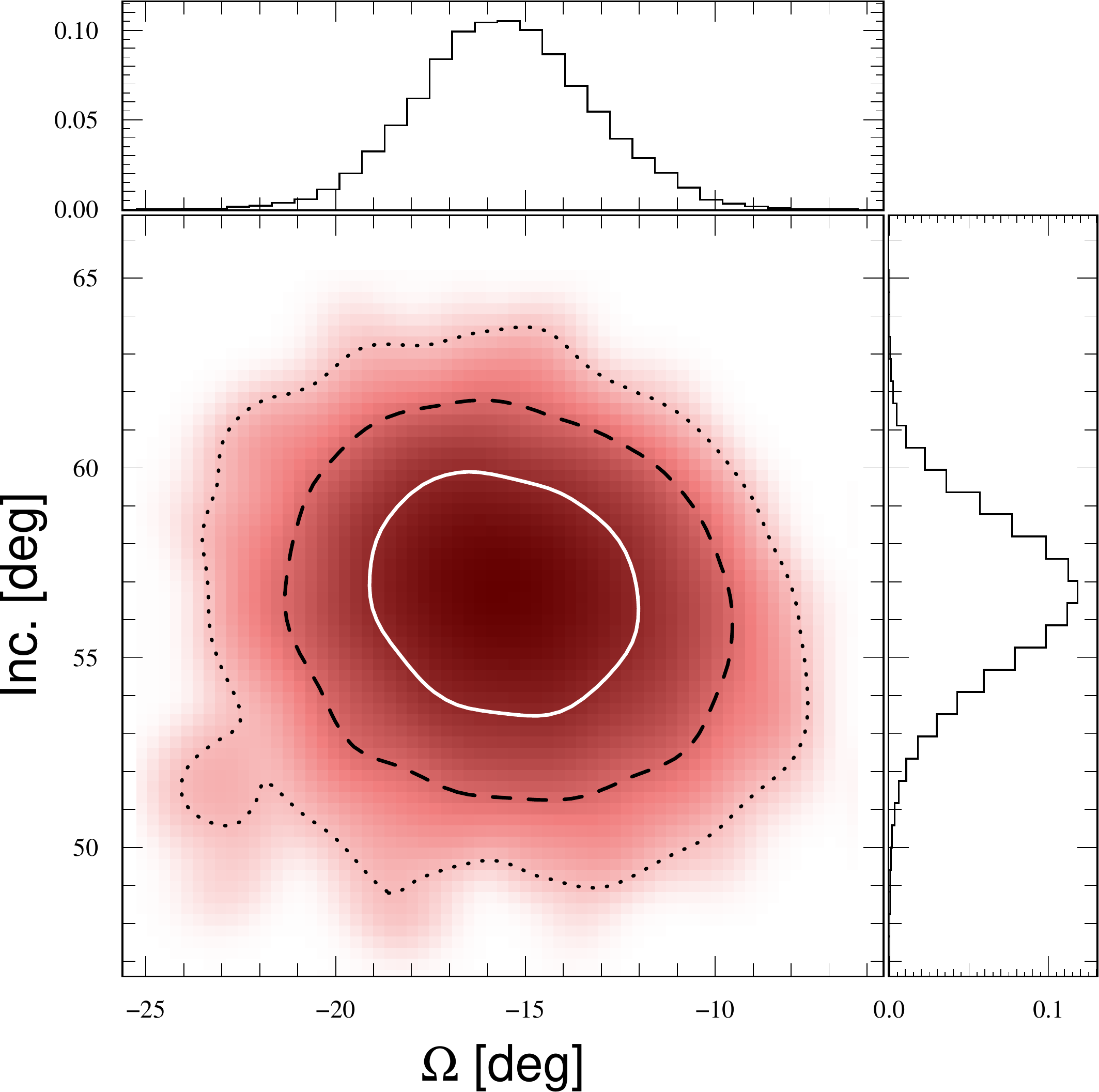}}}
\caption{Joint marginal distribution of several parameters for \dwnine~obtained from $7.5\cdot10^{6}$ MCMC iterations. In each panel, contour lines represent the 1-, 2-, and 3-$\sigma$ confidence intervals of the joint distribution. Sub-panels show the marginal distributions. The top-left panel represents the joint distribution of the parameters $\sqrt{e} sin{(\omega)}$ and$ \sqrt{e} cos{(\omega)}$ with grey circles representing eccentricity values of 0.125, 0.25, 0.5, 0.75, and 1.0.}
\label{fig:PDF2Param}
\end{figure*}
}

\begin{table}[h!]	
\caption{Physical and orbital parameters of the \dwnine~system.}    
\label{table:1}      
\centering                       
\begin{tabular}{l c c c} 
\hline\hline                 %
{\small 2MASS} entry&  & \multicolumn{2}{c}{J08230313--4912012} \\ 
$m_I$ (primary)& (mag)  & \multicolumn{2}{c}{17.1} \\ 
$\Delta\alpha^{\star}_0$& (mas)  & \multicolumn{2}{c}{-137.65$_{   -0.18}^{   +0.20}$}\\  [1pt]
$\Delta\delta_0$ &(mas) &    \multicolumn{2}{c}{-30.72$_{   -0.41}^{   +0.45}$}\\  [1pt]
$\varpi$& (mas) &     \multicolumn{2}{c}{48.09 $\pm$      0.18} \\ [1pt]
$\mu_{\alpha^\star}$ &(mas yr$^{-1}$)   &   \multicolumn{2}{c}{-154.30 $\pm$      0.12} \\  [1pt] 
$\mu_\delta$ &(mas yr$^{-1}$)  &      \multicolumn{2}{c}{7.46 $\pm $     0.09} \\  [1pt] 
$e$ & $\cdots$ &          \multicolumn{2}{c}{0.345$_{   -0.064}^{   +0.068}$} \\  [1pt] 
$\omega$ &(deg) &        \multicolumn{2}{c}{ 36.3$_{  -8.4}^{   +7.2}$}\\  [1pt] 
$P$ &(day) &    \multicolumn{2}{c}{246.36$_{-1.35}^{+1.38}$}\\  [1pt] 
$\lambda_0$ & (deg)&  \multicolumn{2}{c}{201.0$_{ -2.6}^{+  2.5}$} \\[1pt] 
$\Omega$& (deg) &    \multicolumn{2}{c}{-15.6$_{   -2.2}^{   +2.3}$}\\  [1pt] 
$i$&(deg)  &     \multicolumn{2}{c}{56.6$_{   -2.1}^{   +1.9}$}\\  [1pt] 
$a_1$ (barycentre) &(mas) &       \multicolumn{2}{c}{4.61$_{   -0.13}^{   +0.15}$}\\  [1pt] 
$M_2^3/(M_1+M_2)^2$& ($M_J$) &   \multicolumn{2}{c}{2.02$_{   -0.15}^{   +0.18} $}\\ [1pt] 
$\rho$  &(mas)         &    \multicolumn{2}{c}{+20.5$_{   -1.9}^{   +2.0}$} \\  [1pt]
$d$ &(mas)        &     \multicolumn{2}{c}{-24.5$_{   -1.7}^{   +1.7}$}\\  [1pt]
$s_\alpha$,$s_\delta$& (mas)& \multicolumn{2}{c}{$0.16\pm0.07$ / $0.12\pm0.05$}\\[1pt]
Reference date &(MJD)& \multicolumn{2}{c}{55821.933543}\\[3pt]
\multicolumn{4}{c}{Derived and additional parameters}\\[3pt]
$\Delta \varpi$& (mas)  &     \multicolumn{2}{c}{+0.062 $\pm$     0.038}\\ [1pt]
$\varpi_{abs}$ &(mas) & \multicolumn{2}{c}{48.16$_{ -0.19}^{+  0.18}$}\\  [1pt] 
Distance &(pc) &     \multicolumn{2}{c}{20.77 $\pm$ 0.08}\\ [2pt] 
Age& (Gyr) & 1 & 0.6 -- 3\\
$M_1$& ($M_J$) &\!\!\!\!\!\! $78.4\pm{7.8}$& 70.2 -- 82.8\\[1pt] 
$M_2$& ($M_{J}$) & \!\!\!\!${28.5}\pm{1.9}$& 26.7 -- 29.4\\ [1pt] 
$q=M_2/M_1$& $\cdots$&\!\!\!\!\!\!  0.364$_{   -0.017}^{   +0.020}$ &\!\!\!\! 0.355 -- 0.380\\[1pt] 
$\bar a$ (relative orbit) &(AU) &\!\!\!\! 0.36 $\pm$ 0.01&\!\!\!0.348 -- 0.365 \\[3pt]
\multicolumn{2}{l}{Number of epochs / frames}  & \multicolumn{2}{c}{14 / 281}\\
$\sigma_{O - C}$ ({\small RA/DEC})& (mas) & \multicolumn{2}{c}{1.05 / 0.82}\\[1pt]
$\sigma_{O - C, Epoch}$ & (mas)  & \multicolumn{2}{c}{0.330}\\ 
$\sigma_{O - C, Epoch}$ ({\small RA})& (mas)  & \multicolumn{2}{c}{0.410}\\ 
$\sigma_{O - C, Epoch}$ ({\small DEC})& (mas)  & \multicolumn{2}{c}{0.221}\\ 
\hline
\end{tabular}
\tablefoot{Parameter values correspond to the median of the marginal parameter distributions and uncertainties represent 1$\sigma$ ranges. The proper motions are not absolute and were measured relative to the local reference frame. The modified Julian date (MJD) equals barycentric Julian date -- 2400000.5. {The semimajor axis of the relative orbit $\bar a$ is given in AU.} The inclination is with respect to the plane of the sky. There is an ambiguity of 180\degr~in inclination and longitude of ascending node that is inherent to astrometric orbits but does not influence the system's properties. Parameters that depend on the age of the system are given with formal uncertainties for an age of 1 Gyr and with value ranges for ages of 0.6 -- 3 Gyr.}
\end{table}

\subsection{Parallax correction}\label{sec:parcor}
Because the target's motion is measured relative to reference stars that are not located at infinite distance, a correction term has to be applied to convert relative parallax $\varpi$ to absolute parallax $\varpi_{abs}$. There are three strategies to perform the parallax correction: (1) Tie the reference frame to extragalactic and therefore quasi-absolute references; (2) Determine photometric distances to the reference stars \citep{Vrba:2004zr, Faherty:2012uq}; (3) Use a Galactic model to estimate the distances of field stars \citep{Andrei:2011lr, Dupuy:2012fk}. Ideally, the parallax correction uncertainty should be smaller than the parallax precision, that is in our case $<0.1\,$mas. Inspection of the target field revealed that extragalactic sources that could be securely identified were extended objects and their photocentres can therefore not be measured with sufficient accuracy. The first method is thus not applicable.
The second method relies on obtaining $V$-$K$ colours of reference stars, for instance those included in the NOMAD catalogue \citep{Zacharias:2004nat}. For \dwnine, the number of usable stars is restricted to ten, which led to poor results and we discarded the second method, too.\\ 
For the third method, we used a model of the Galaxy \citep{Robin:2003fk} to obtain a synthetic sample of stars in a $10\arcmin\times10\arcmin$ field centred on the target and selected the magnitude range covered by the reference stars used for determining the parallax. The measured magnitude distribution is well matched by the model (Fig.~\ref{fig:7},a). In our model adjustments, negative parallaxes are allowed and thus the measured parallax distribution appears shifted towards smaller values relative to the model parallax distribution. The average value of this shift corresponds to the parallax correction. We discarded the 10th percentile of largest and smallest parallaxes, i.e.\ only data located between the two dashed lines in Fig.~\ref{fig:7},b were considered, and obtained a parallax correction of $\Delta \varpi = +0.062 \pm 0.038$\,mas (s.e.m. using 283 stars) to be added to the relative parallax of \dwnine. Its value is small and comparable to its uncertainty, which is expected because the reference stars are very faint (17th - 22nd magnitude in $I$-band) and therefore located at large distances. The correction corresponds to an average distance of the reference stars of $\sim$16~kpc, which is compatible with the extent of the Galaxy (\dwnine~is located towards the Vela constellation). We have tested the Galactic model method with a brighter target of our astrometric survey, for which 21 reference stars could be used to apply the photometric distance method. The results of both methods agreed, thus validating our approach.

\subsection{Primary mass estimation}\label{sec:massestim}
We cannot measure the mass of \dwnine~directly and therefore have to rely on an indirect estimate obtained from spectro-photometric measurements, the absolute parallax, and an age estimate in combination with evolutionary models of substellar and stellar objects. We compared the optical spectrum of \dwnine~\citep{Phan-Bao:2008fr} to spectral standards from \cite{Martin:1999yf} and found that the gravity-sensitive {Na\,I} equivalent width {of $3.4\! \pm \!0.5\,\AA$} and the TiO band around 8400\,\AA~indicate that the object is intermediate between the 'old' L1 spectral standard DENIS-P J1441-0945 and the 'young' L1 standard G 196-3 B (Fig.~\ref{fig:spectrum}, {see also Fig.\,3 of }\citealt{Martin:2010qy}). \dwnine~shows a hint of Li absorption but the low signal-to-noise prevents us from claiming a detection on the basis of this spectrum alone. We therefore adopted an age range of 0.6 - 3 billion years (Gyr) for \dwnine. The tangential velocity of 15 km\,s$^{-1}$ for \dwnine~is comparable to the values of M7-L2 dwarfs in the field \citep{Schmidt:2007ve}, having an estimated age range of 2--4 Gyr.
\dwnine~is a photometrically calm object with no noticeable activity. We monitored its brightness relative to field stars over the observation timespan and found that photometric fluctuations do not exceed the typical measurement uncertainty of $\pm 0.004$ magnitudes. Optical and infrared photometric measurements of \dwnine~were retrieved from the catalogues {\small 2MASS} (\citealt{Skrutskie:2006fk}; $J$,$H$,$K$ bands) and {\small DENIS} (\citealt{Phan-Bao:2008fr}; $I$,$H$,$K$ bands). Absolute magnitudes were obtained on the basis of the measured absolute parallax. The BT-Settl models \citep{Chabrier:2000kx,Allard:2012uq} were linearly interpolated in mass for a given age to determine the best-fit mass using a least-square minimisation taking into account the magnitude and parallax uncertainties. The formal errors of this procedure were very small ($<0.5\%$) and we adopted a constant 10~\% mass uncertainty instead to account for potential model inaccuracies (see e.g.\ \citealt{Dupuy:2009fr}). An alternative mass estimation method using the bolometric luminosity is discussed in Appendix~\ref{sec:bol} and yields similar results. The mass estimate of \dwnine~at 1 Gyr is $0.075\pm 0.007\,M_{\sun}$ and in the age range of $0.6-10$ Gyr the corresponding mass lies in the range of $0.067-0.079 \,M_{\sun}$ {with essentially no mass variation for ages older than 3 Gyr}. Follow-up spectroscopy is required to confirm the youth indicators of \dwnine~and to obtain a refined age estimate.
\begin{figure}[h!]
\centering
\includegraphics[width = 0.9\linewidth]{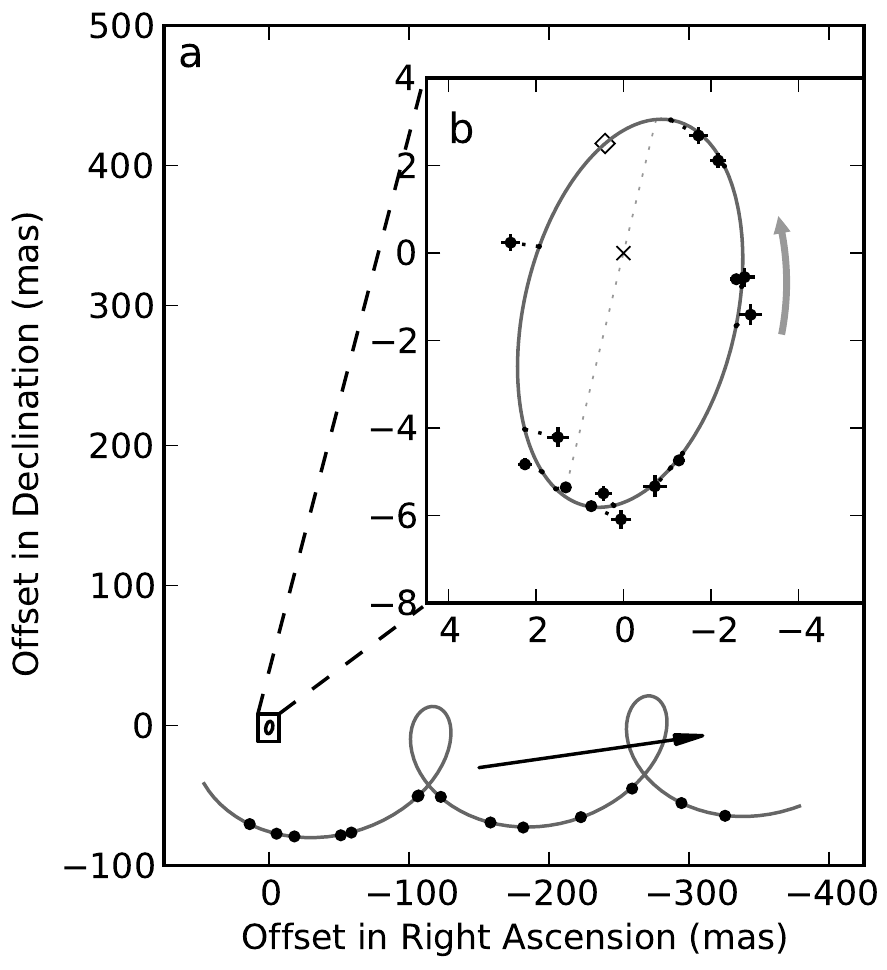}
\caption{Astrometric motion of \dwnine~and its barycentric orbit. Panel a shows proper and parallactic motion relative to the field of reference stars. The astrometric observations and the model are shown as black circles and grey curve, respectively. The black arrow indicates the direction and amplitude of the proper motion over one year. Panel b is a close-up of the barycentric orbit caused by the gravitational pull of the orbiting brown dwarf. Observations with s.e.m. error bars and the best-fit model are shown as black circles and grey curve, respectively. The barycentre and periastron position are marked with a cross and an open square, respectively, and the dotted line represents the line of nodes.} 
\label{fig:orbit}
\end{figure}
\section{Orbit and companion mass of \dwnine}
Proper and parallactic motion of the target are of the order of hundreds and tens of mas, respectively (Fig.~\ref{fig:orbit}a), and are superimposed on the orbital motion with an amplitude of several mas (Fig.~\ref{fig:orbit}b). The parallax determines the target's distance, thus allowing us to convert angular measurements to linear quantities, and the orbital parameters yield information on the physical properties of the binary system. 
For the estimated age range of $0.6-3$ Gyr, \dwnine~has a mass in the range of $0.067-0.079\,M_{\sun}$, thus encompassing the theoretical hydrogen burning mass limit of $\sim\!\!0.075\,M_{\sun}$ assuming Solar metallicity. \dwnine~is therefore either a very low-mass star or a brown dwarf, an ambiguity that leaves our main finding unaffected.\\
The parallactic motion of \dwnine~reveals the system's distance of $20.77 \pm 0.08$ parsec from Earth, in agreement with an earlier photometric estimate by \cite{Phan-Bao:2008fr} but determined with a relative precision of 0.4~\%. Our measurements show that \dwnine~moves on a {photocentric} orbit with a semimajor axis of $4.61_{   -0.13}^{   +0.15}$ mas and a period of $246.4\pm1.4$ days (Table~\ref{table:1}). The orbit is eccentric ($0.35_{   -0.06}^{   +0.07}$) and is observed with an inclination of $56.6_{   -2.0}^{   +1.9}$\degr (Fig.~\ref{fig:orbit}b). {Using the photocentric semimajor axis as an approximation of the barycentric orbit size, we estimated the secondary mass and the $I$-band magnitude difference between primary and secondary of $\Delta m_I = 5.1 - 8.4$ in the $0.6-3$ Gyr age range. We therefore assumed that the companion's light contribution is negligible and that the barycentric orbit coincides with the photocentric orbit.} The system's mass ratio lies in the range of 0.355--0.380 and is determined by the semimajor axis and the primary mass estimate, where the age uncertainty dominates the mass ratio uncertainty. The corresponding companion mass is 26.7--29.4$\,M_{J}$ and falls into the mass range of giant planets and low-mass brown dwarf companions around Sun-like stars \citep{Sahlmann:2011fk}. Figure 2 shows an overview of known ultracool binaries and the particular location occupied by \dwnine~having a relative separation of $0.36\pm0.01$ AU.
\begin{figure*}
\sidecaption
\includegraphics[width = 6cm]{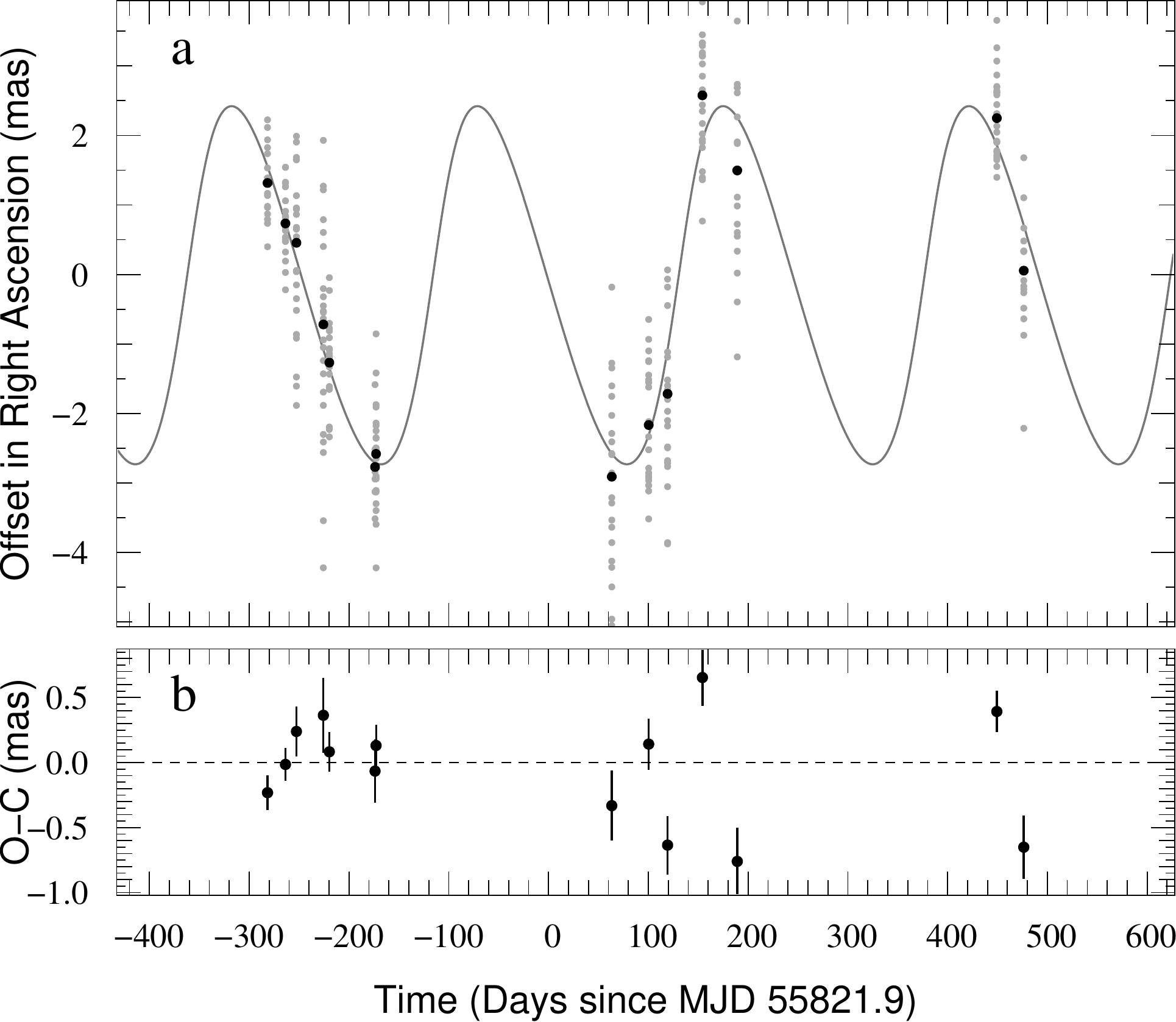}\hspace{3mm}
\includegraphics[width = 6cm]{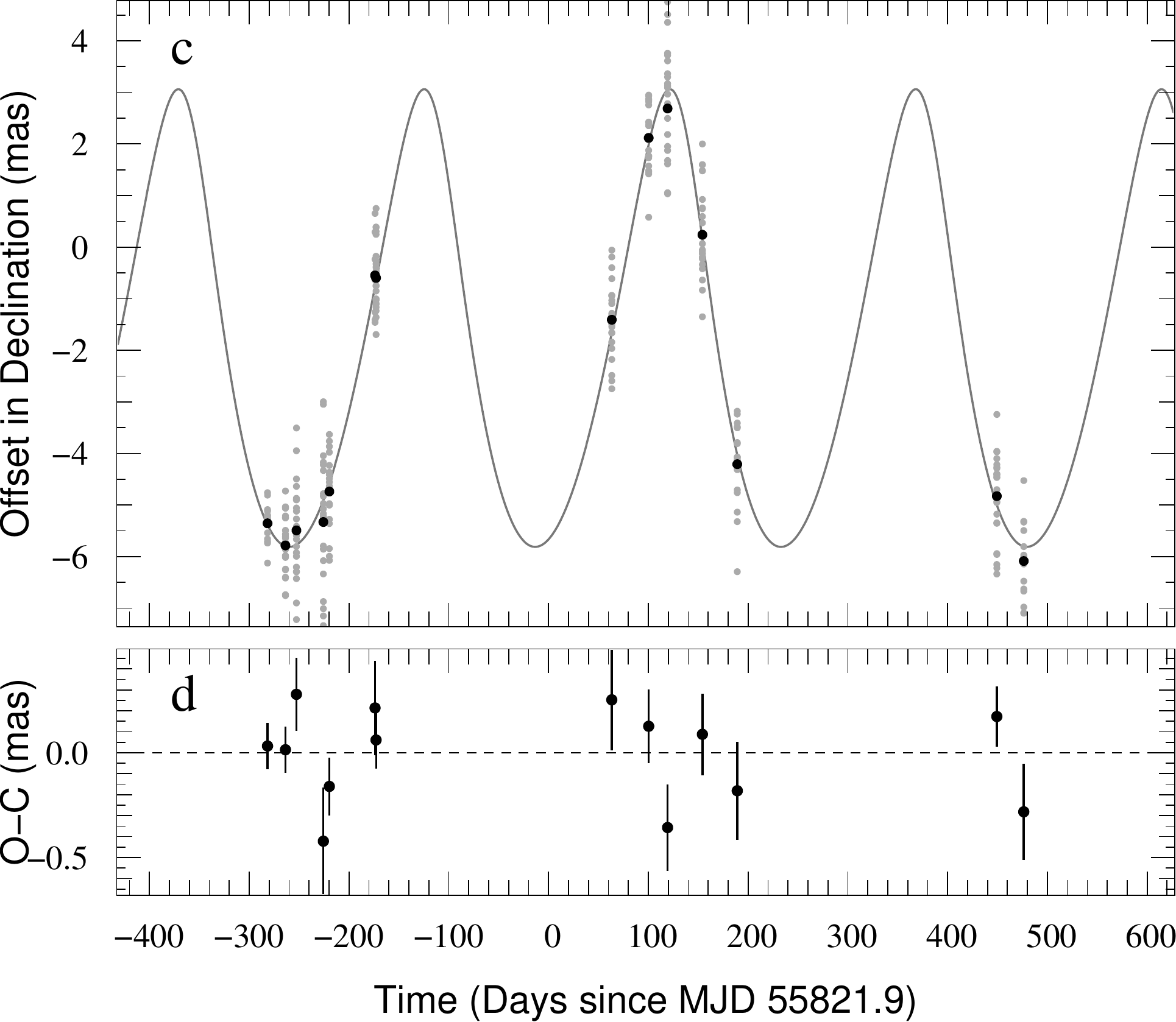}
\caption{Orbital motion of \dwnine~as a function of time. The orbital signature in right ascension (panel a) and declination (panel c) is shown, where black symbols show epoch average values and grey symbols indicate the individual frame measurements. Panels b and d show the observed minus calculated residuals of epoch averages.}
\label{fig:axt}
\end{figure*}

\section{Discussion and conclusions}
The discovery of the \dwnine~system is {unusual} in the context of the known ultracool binary population {because it has a particularly small mass ratio and is located in a {sparsely} populated region of the separation -- mass-ratio plane shown in Fig.~\ref{fig:2}}. Ultracool binaries were previously found preferentially in nearly equal mass configurations ($q\gtrsim0.7$) and with a separation distribution peaked at $\sim$1-10 AU (Fig.~\ref{fig:2}), in contrast to Solar type and M dwarf binaries that show a flat mass-ratio distribution for $q\gtrsim0.2$ and a separation distribution peaked at $\sim$\,25-35 AU \citep{Raghavan:2010fk,Janson:2012ly}. The sharp decline in the number of ultracool binaries with separations $\lesssim$ 1 AU coincides with the typical resolution limit of current telescopes that impedes detection of these binaries with the most successful method of direct imaging {that is most sensitive to systems with nearly equal masses because of the favourable brightness contrast}. Radial velocity observations are sensitive to small separation binaries but merely yield a lower limit to the mass ratio assuming the primary mass can be estimated, in particular for low mass-ratio systems that constitute single-lined spectroscopic binaries \citep{Joergens:2007nx, Blake:2008fr}. {Observations aimed at detecting unresolved binaries through differences in the components' spectra \citep{Burgasser:2010kx} may identify binaries like \dwnine.} {Gravitational microlensing events were used to discover two binaries {seen} close to \dwnine~in Fig.~\ref{fig:2} \citep{Choi:2013fk}. Those systems are located at large distances from Earth (400 and 2000 parsec) and have small total masses (0.025 and 0.034 $M_\sun$) compared to \dwnine~having $M_1+M_2\simeq0.10\,M_\sun$. Microlensing events typically yield unrepeatable snapshots and therefore the eccentricities and orbital periods of those binaries will remain unknown. However, the recent discovery of these three systems indicate that small-separation binaries with small mass-ratios may not be as rare as previously thought.
\begin{figure}[h!]
\centering
\includegraphics[width = \linewidth]{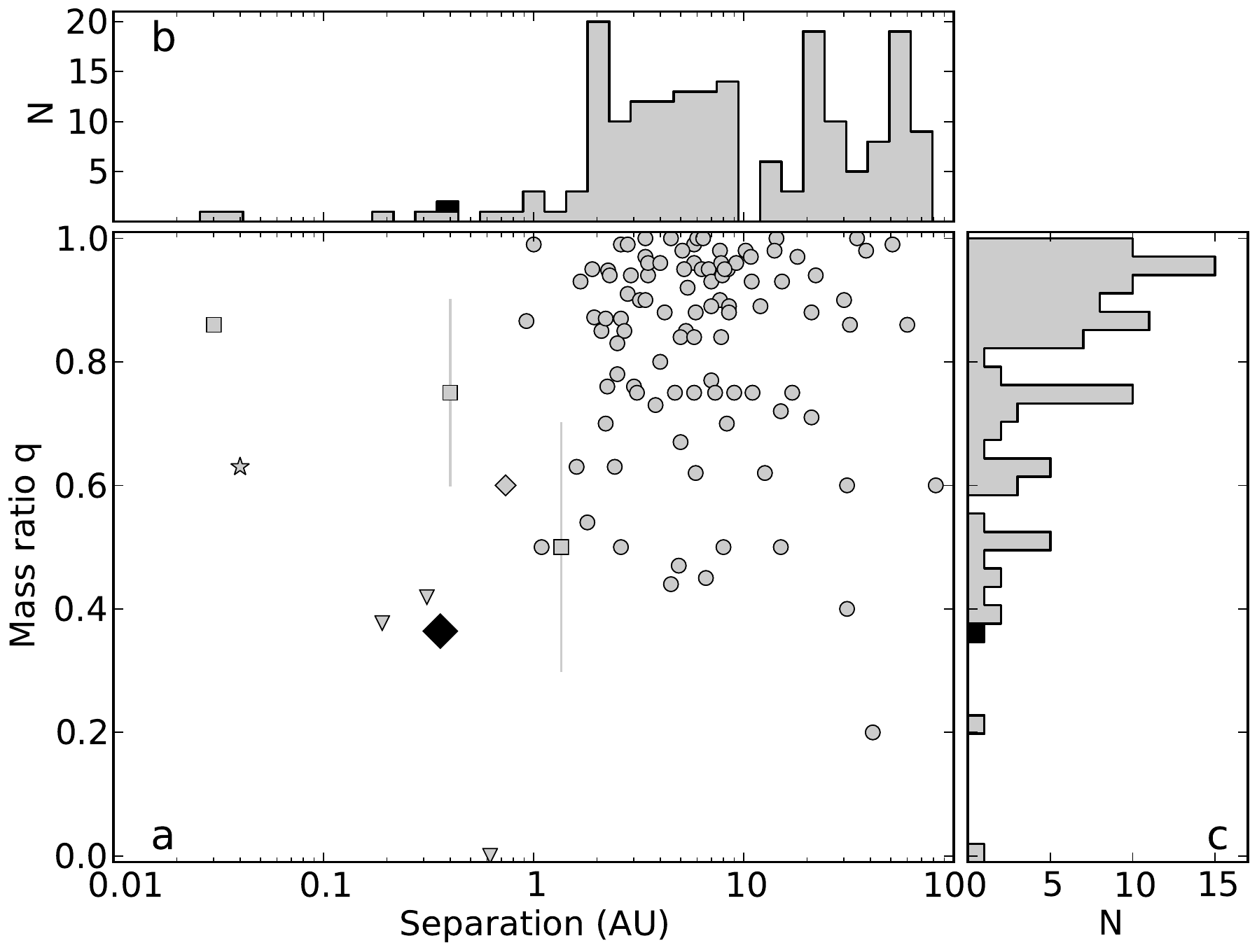}
\caption{Separations and mass ratios of very low-mass binaries. Panel a shows the mass ratio as a function of separation for 108 very low mass binaries ($M_1+M_2 < 0.2\,M_{\sun}$, Sect.~\ref{sec:vlmb}). The location of the \dwnine~binary is indicated with a black diamond and uncertainties are smaller than the symbol size. Binaries with separations $<$ 100 AU are shown, where the separation is either given by the the orbit's semimajor axis or, in most cases, the actual observed projected separation of directly imaged binaries. Error bars are not shown for clarity of display. Symbols indicate the respective detection method: diamonds, astrometry; squares, radial velocity; triangles, gravitational microlensing; star, eclipse photometry; circles, direct imaging or spectroscopy. For radial velocity systems with unknown mass ratio, the range of possible values is indicated. Panel b shows the distribution in separation. The two smallest separation binaries are a short period spectroscopic binary \citep{Basri:1999kx} and an eclipsing binary \citep{Stassun:2006pt}. Panel c shows the distribution in mass-ratio. The two systems with mass ratios smaller than \dwnine~are a planetary mass companion detected through microlensing \citep{Kubas:2010fk} and a directly imaged companion to a young brown dwarf \citep{Chauvin:2004ve}.} 
\label{fig:2}
\end{figure} 

With our observational procedures and reduction methods we have demonstrated the capability of ground-based optical astrometry to achieve {200} $\mu$as astrometry on faint optical sources over {a large field of view (a few arcminutes)}. For ultracool dwarfs, the astrometric performance of {\small FORS2} is therefore comparable to what is expected from the Gaia space astrometry mission (e.g.\ \citealt{2011AdSpR..47..356M}). This opens a new window to the parameter space of small mass ratios and {small-to-intermediate} separations of ultracool binaries. Astrometric surveys will contribute to the comprehensive characterisation of ultracool binaries by measuring their frequency at {separations of }$\sim\,$0.1--10~AU, estimated in the range of 1--30 \% \citep{Guenther:2003kx, Basri:2006fk, Joergens:2008qy, Blake:2010lr}, and by {refining} their eccentricity distribution. The observational evidence will help resolving the question whether ultracool binaries form like stellar binaries \citep{Thies:2008ve, Parker:2011qf}. In addition to its sensitivity to companions, astrometry yields a direct measurement of the target's distance from the Earth that is essential to understand the physics of ultracool dwarfs. The parallax accuracy achieved here allows distance determinations at an unprecedented precision for ground-based optical astrometry, thus removing distances as a dominant source of uncertainty in the modelling of ultracool dwarfs. Finally, the astrometric performance demonstrated here is sufficient to discover planetary companions of nearby ultracool dwarfs with masses as low as one Neptune mass at separations $\gtrsim\,$0.4 AU. 

\begin{acknowledgements}
J.S., D.S., M.M., D.Q., and S.U. thank the Swiss National Science Foundation for supporting this research. E.M. was supported by the Spanish Ministerio de Economia y Competitividad through grant AyA2011-30147-C03-03 and thanks the Geosciences Department at the University of Florida for a visiting appointment. J.S. thanks T. Dupuy for sharing his VLM binary compilation and kindly acknowledges support as a visitor at the Centro de Astrobiolog\'ia in Villanueva de la Ca\~nada (Madrid). We thank N. Phan-Bao for making the spectrum of \dwnine~available to us. We thank the ESO staff for efficiently scheduling and executing our observations. This publication makes use of the Very-Low-Mass Binaries Archive housed at http://www.vlmbinaries.org and of data products from the Two Micron All Sky Survey, which is a joint project of the University of Massachusetts and the Infrared Processing and Analysis Center/California Institute of Technology, funded by the National Aeronautics and Space Administration and the National Science Foundation.
\end{acknowledgements}

\bibliographystyle{aa} 
\bibliography{/Users/sahlmann/astro/papers} 

\begin{appendix}
\section{Data covariance}\label{sec:cov}
Due to the reduction procedure, the individual frame measurements within one epoch are not independent but are subject to correlated noise of equal magnitude in {\small RA} and {\small DEC}. Therefore, the corresponding covariance matrix contains non-zero off-diagonal elements and has a block-diagonal form with the nominal uncertainties on the diagonal and the covariance amplitude on the off-diagonal elements for each epoch. The covariance between data taken at different epochs is always zero. Because consideration of the covariances is expensive in computation time, we neglect the off-diagonal terms of the covariance matrix to run the principal part of the {\small MCMC}. We quantified the effect of considering the full covariance matrix by running a {\small MCMC} with a smaller number ($10^5$) of iterations. The results in both cases are statistically indistinguishable because the  parameter standard deviations are equivalent and the differences in the median parameter values are smaller than $\sigma$/10. The only significant difference is a smaller resulting nuisance parameter in {\small RA} ($s_\alpha$) when considering the full covariance matrix (0.08 mas compared to 0.16 mas with a diagonal covariance matrix), indicating that this parameter sensibly accounts for additional signals in the form of correlated noise. We conclude that the off-diagonal terms of the covariance matrix can be neglected when employing the analysis methods presented {here}.

\section{Comparison with Galaxy model}
Figure~\ref{fig:7} illustrates how we use a model of the Galaxy to determine the parallax correction $\Delta \varpi$.
\begin{figure}[h!]
\centering
\includegraphics[width=0.8\linewidth]{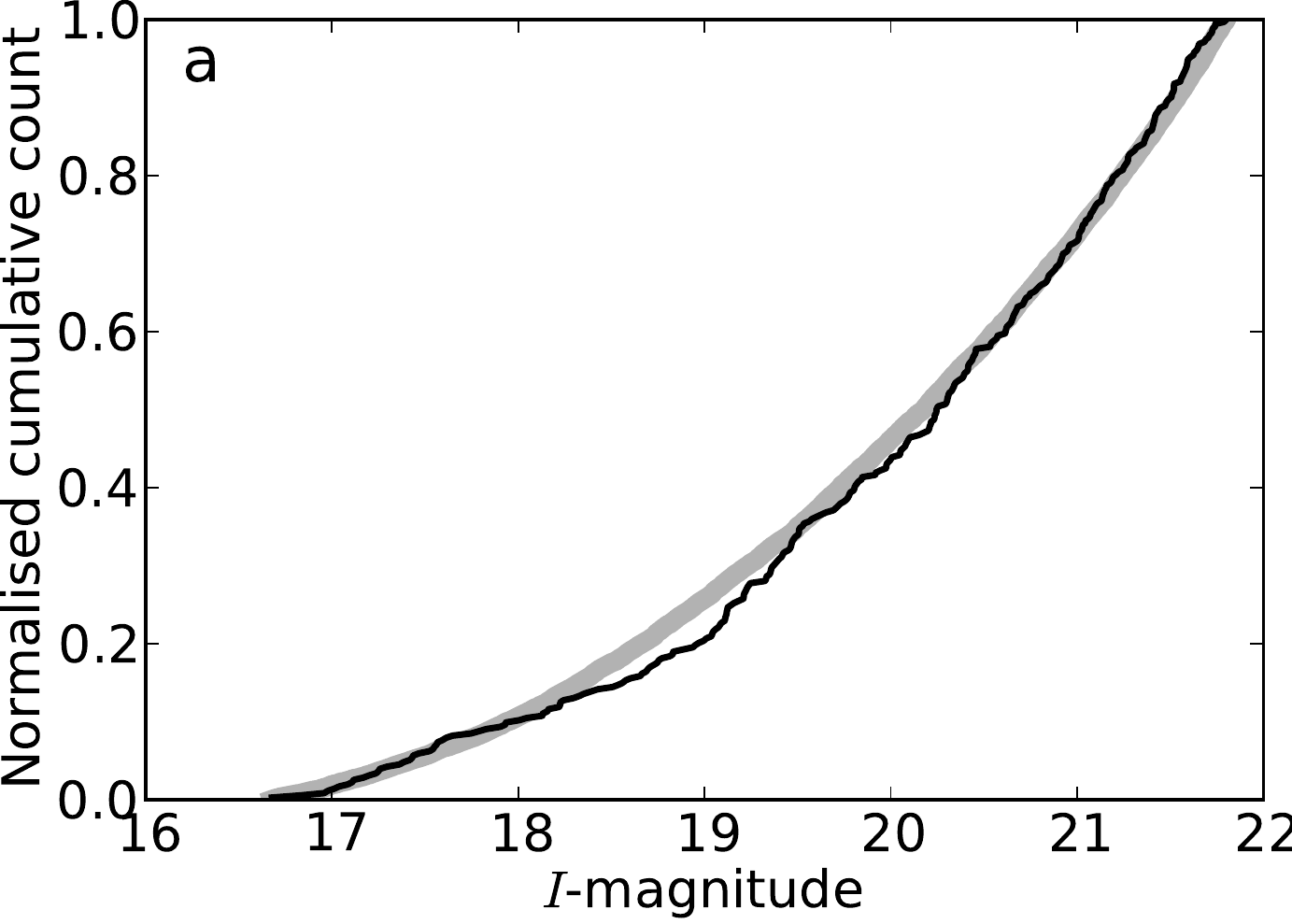}\hspace{2mm}
\includegraphics[width=0.8\linewidth]{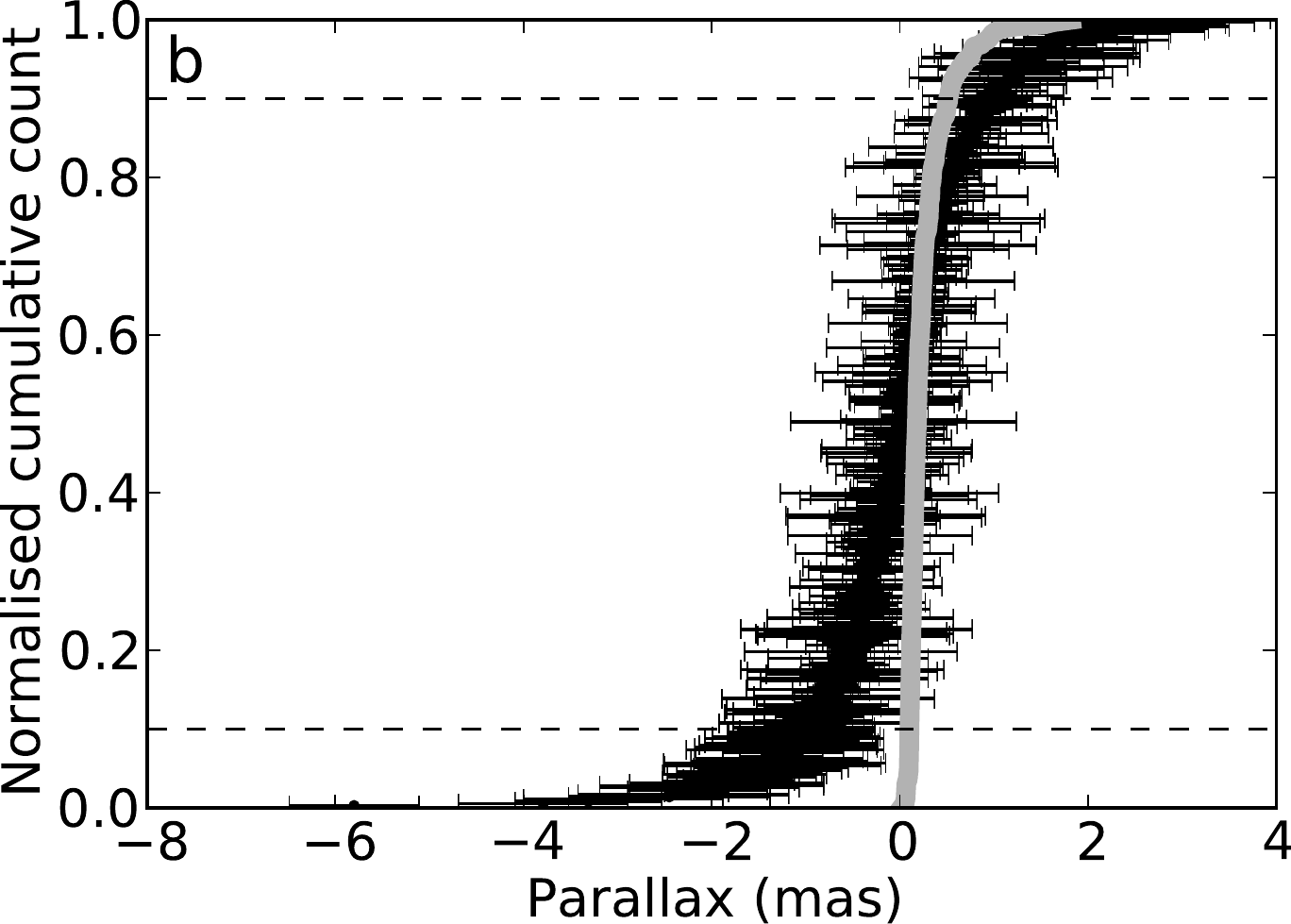}
\caption{Using a galaxy model to determine the parallax correction. Cumulative distribution of magnitudes (panel {a}) and parallaxes (panel {b}) for the 283 reference stars of \dwnine. The model and measured data are shown in grey and black, respectively.}
\label{fig:7}
\end{figure}

\section{Primary mass estimation using bolometric luminosity}\label{sec:bol}
{As a consistency check, we used the method of Sect.~\ref{sec:massestim} restricted to the three {\small 2MASS} bands and compared the results to a method relying on an estimation of the bolometric luminosity. We converted {\small 2MASS} magnitudes to the {\small MKO} system using updated colour transformations \citep{Carpenter:2001ys}\footnote{\url{http://www.astro.caltech.edu/~jmc/2mass/v3/transformations/}} and bolometric corrections \citep{Liu:2010fk} to obtain the luminosity and we assumed an uncertainty of one spectral type subclass. The corresponding mass at a given age was found by interpolating the BT-Settl models. Differences between $J$,$H$,$K$ are negligible and we used their average. The resulting masses lie a few percent higher than the estimation with the previous method but both methods yield compatible results within the {adopted} 10\% uncertainty. At 1 Gyr the alternative mass estimate is $0.079\pm0.001$, where the error reflects only the uncertainty in spectral type. In the age range of 0.6 -- 10 Gyr, the alternative mass range is $0.072 - 0.081 \,M_{\sun}$.  }
\begin{figure}[h!]
\centering
\includegraphics[width = \linewidth,trim = 0cm 0.7cm 0cm 1cm, clip]{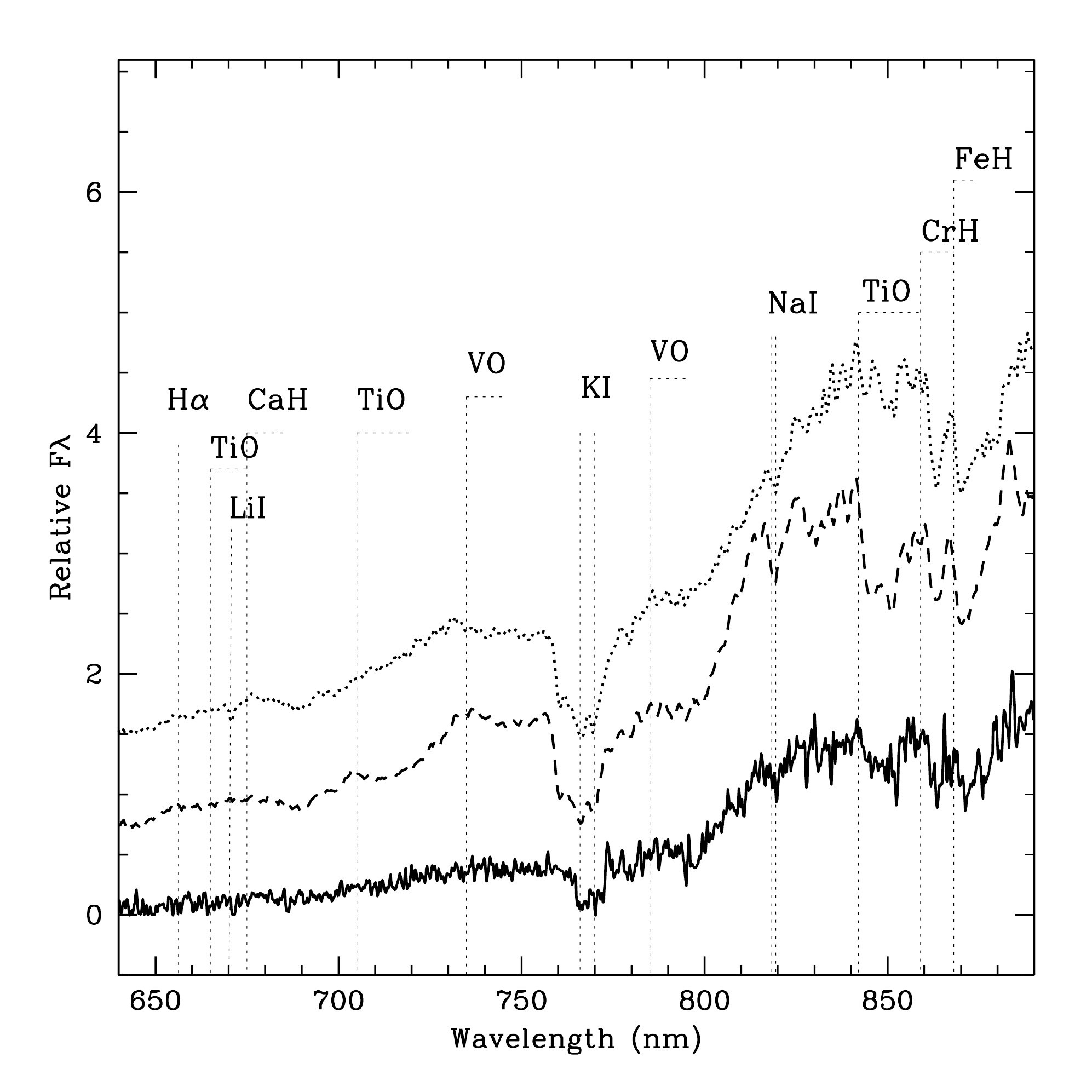}
\caption{Comparison of the optical spectra of \dwnine~from \cite{Phan-Bao:2008fr} (solid), the 'old' L1 dwarf DENIS-P J1441-0945 (dashed), and the 'young' L1 dwarf G 196-3 B (dotted), both from \cite{Martin:1999yf}. The relevant spectral features are labelled.}
\label{fig:spectrum}
\end{figure}

\section{Compilation of very low mass binary systems}\label{sec:vlmb}
The sample of binaries shown in Fig. \ref{fig:2} was constructed on the basis of the compilation at \url{vlmbinaries.org}. Because its last update was in July 2009, we searched the literature for new systems and revised parameters. We added 24 systems from \cite{Choi:2013fk, Burgasser:2011lr,Allers:2010ly,Gelino:2010fk,Liu:2010fk,Liu:2011rt,Kraus:2012dq,Gelino:2011yq,Chauvin:2012fk,Geyer:1988kx,Phan-Bao:2006fk,Allers:2009bh,Burgasser:2009hb} and we updated system parameters when applicable using the compilations of \cite{Liu:2010fk,Dupuy:2011uq,Kraus:2012dq}.
\end{appendix}
\end{document}